\documentclass[reqno,oneside,12pt]{amsart}

\usepackage{amssymb,graphicx,braket,amsmath,amsfonts,amsthm}
\usepackage{hyperref}
\usepackage[utf8]{inputenc}
\usepackage{braket}
\usepackage{color}
\usepackage{amsaddr}
\usepackage{tikz}
\usetikzlibrary{shapes,arrows,er,positioning}

\theoremstyle{plain}
\newtheorem{theorem}{Theorem}[section]
\theoremstyle{plain}
\newtheorem{proposition}[theorem]{Proposition}
\newtheorem*{proposition*}{Proposition}
\newtheorem{remark}[theorem]{Remark}
\newtheorem*{theorem*}{Theorem}

\newtheorem*{assumption*}{Assumption}
\newtheorem{corollary}[theorem]{Corollary}

\theoremstyle{definition} 
\newtheorem{definition}[theorem]{Definition}

\numberwithin{equation}{section}

\def\mainmatter{\def\baselinestretch{1.1}\normalfont}

\usepackage[margin=1in]{geometry}

\makeatletter
\renewcommand{\section}{\@startsection
{section}
{1}
{\z@}
{-\baselineskip}
{0.8\baselineskip}
{\centering\scshape\large}} 

\renewcommand{\subsection}{\@startsection
{subsection}
{2}
{\z@}
{-0.8\baselineskip}
{0.5\baselineskip}
{\normalfont \bf \normalsize}} 

\renewcommand{\subsubsection}{\@startsection
{subsubsection}
{3}
{\z@}
{-0.8\baselineskip}
{0.5\baselineskip}
{\normalfont \it \normalsize}} 
\makeatother

\newcommand{\arxiv}[1]{\href{http://arxiv.org/abs/#1}{\texttt{arXiv\string:\allowbreak#1}}}

\begin{document}

\title{$\mathcal{N}=1$ Super Topological Recursion}

\author{Vincent Bouchard}
\address{Department of Mathematics and Statistical Sciences,
University of Alberta, 632 CAB,
Edmonton, Alberta, T6G 2G1, Canada}
\email{vincent.bouchard@ualberta.ca}

\author{Kento Osuga}

\address{School of Mathematics and Statistics,
University of Sheffield,
The Hicks Building,
Hounsfield Road,
Sheffield, S3 7RH,
United Kingdom}
\email{k.osuga@sheffield.ac.uk}

\begin{abstract}

	We introduce the notion of $\mathcal{N}=1$ abstract super loop equations, and provide two equivalent ways of solving them. The first approach is a recursive formalism that can be thought of as a supersymmetric generalization of the Eynard-Orantin topological recursion, based on the geometry of a local super spectral curve. The second approach is based on the framework of super Airy structures. The resulting recursive formalism can be applied to compute correlation functions for a variety of examples related to 2d supergarvity.

\end{abstract}

\newpage
\maketitle

\setcounter{tocdepth}{1}
\tableofcontents
\mainmatter

\newpage
\section{Introduction}\label{sec:intro}
The Eynard-Orantin topological recursion introduced in \cite{CEO,EO,EO2} can be used to compute various kinds of enumerative invariants, such as Gromov-Witten invariants, Hurwitz numbers, knot invariants, and more (see \cite{BKMP,BM,BEM,DBOSS,EMS,EO3,FLZ2,FLZ3,GJKS,Ma} and references therein). Starting with a spectral curve, the Eynard-Orantin topological recursion provides an infinite sequence of multilinear differentials (known as correlation functions) which are generating functions for those enumerative invariants. 

The topological recursion does not come out of nowhere. It can be obtained as a unique solution (respecting polarization) of a set of equations known as abstract loop equations, which were formalized in \cite{BEO}. (The well-known loop equations for Hermitian matrix models fit into this abstract framework.) Concretely, the Eynard-Orantin topological recursion solves the loop equations through residue analysis at the poles of the correlation functions.

Recently, Kontsevich and Soibelman developed the framework of Airy structures \cite{KS, ABCD}. The concept of Airy structures can be thought of as an algebraic reformulation (and generalization) of the Eynard-Orantin topological recursion. Given the data of a spectral curve, one can construct a corresponding Airy structure, and its associated partition function contains the same information as the correlation functions of the Eynard-Orantin topological recursion. In fact, as explained in \cite{ABCD,HAS}, one can think of Airy structures as providing another approach to solving abstract loop equations. Namely, the abstract loop equations can be transformed into a set of differential constraints satisfied by a partition function. These differential operators satisfy the defining properties of an Airy structure, and hence the resulting partition function is uniquely defined by the differential constraints.\footnote{\cite{HAS} discusses the more general equivalence between higher abstract loop equations, the Bouchard-Eynard topological recursion, and higher Airy structures.} Moreover, in the simple context of a local spectral curve with one component, these differential operators form a (suitably polarized) representation of the Virasoro algebra. In this way, the abstract loop equations are reformulated as Virasoro constraints, and the framework of Airy structures guarantees that these Virasoro constraints have a unique solution.


Schematically, one could summarize the relations among abstract loop equations, the Eynard-Orantin topological recursion, and Airy structures as follows:

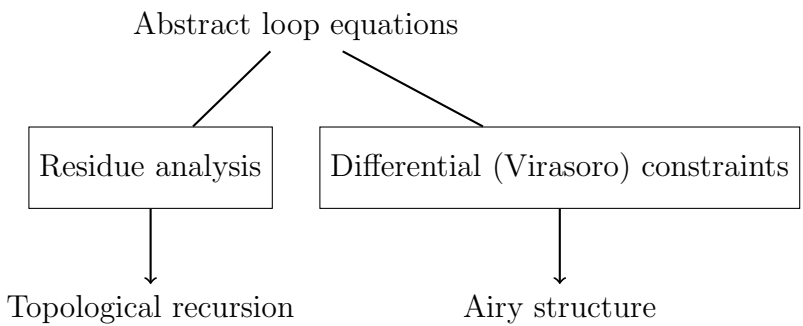
\begin{figure}[h]
\begin{tikzpicture}
\node (eq) {Abstract loop equations};
\node[entity, below left= 1cm and -2cm of eq] (P) {Residue analysis};
\node[entity, below right= 1cm and -2cm of eq] (V) {Differential (Virasoro) constraints};
\node[below=1cm of P] (TR) {Topological recursion};
\node[below=1cm of V] (AS) {Airy structure};
\draw[-,draw=black, thick] (eq) to (P);
\draw[-,draw=black, thick] (eq) to (V);
\draw[->,draw=black,thick] (P) to (TR);
\draw[->,draw=black,thick] (V) to (AS);
\end{tikzpicture}
\caption{Two dual ways of solving abstract loop equations.}
\end{figure}

Supersymmetric generalizations of the Eynard-Orantin topological recursion have been discussed in \cite{C1,C2,C,BO,O}
in the context of supereigenvalue models.
On the other hand, from an algebraic point of view, a supersymmetric generalization of Airy structures (super Airy structures) was proposed in \cite{SAS}, with a corresponding existence and uniqueness theorem for the associated partition function. 
However, the relation between these two approaches is not obvious. Furthermore, it is not clear what a natural supersymmetric generalization of the Eynard-Orantin topological recursion should look like, which would play the role of a ``dual'' to super Airy structures.


The goal of this paper is to fill the gap. Our approach is to start with the notion of $\mathcal{N}=1$ abstract super loop equations. We define a natural notion of super loop equations, as a generalization of the standard loop equations. Then, through residue analysis, we show that if a solution of these super loop equations that respects the polarization exists, it must be constructed recursively by what we call the ``$\mathcal{N}=1$ super topological recursion'', which provides a generalization of the Eynard-Orantin topological recursion. The initial data is formulated in terms of a local super spectral curve. Second, we show that the abstract super loop equations can also be transformed into differential constraints, which take the form of a super Airy structure. The unique associated partition function then reconstructs the solution of the abstract super loop equations, and the framework of super Airy structure guarantees its existence and uniqueness. Furthermore, in the context of a local super spectral curve with one component (which is what we mainly focus on in this paper), these differential operators form a (suitably polarized) representation of the $\mathcal{N}=1$ super Virasoro algebra in the Neveu-Schwarz sector. We have thus reformulated the abstract super loop equations as super Virasoro constraints, and the framework of super Airy structures guarantees that these super Virasoro constraints have a unique solution.

This is encapsulated in the following figure:

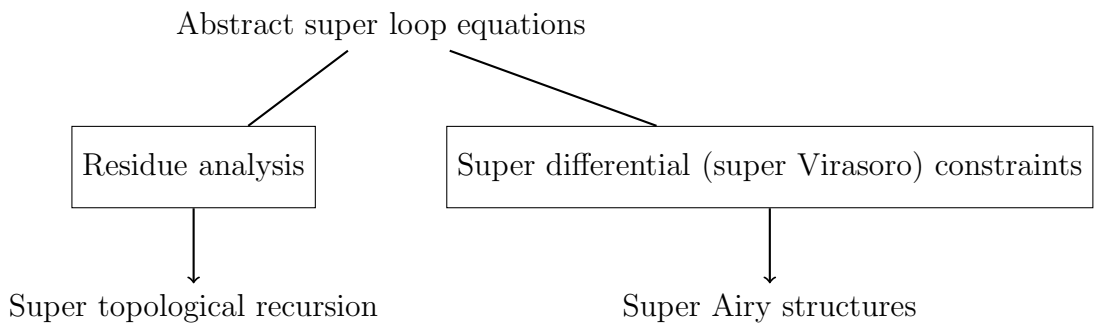
\begin{figure}[h]
\begin{tikzpicture}
\node (eq) {Abstract super loop equations};
\node[entity, below left= 1cm and -2cm of eq] (P) {Residue analysis};
\node[entity, below right= 1cm and -2cm of eq] (V) {Super differential (super Virasoro) constraints};
\node[below=1cm of P] (TR) {Super topological recursion};
\node[below=1cm of V] (AS) {Super Airy structures};
\draw[-,draw=black, thick] (eq) to (P);
\draw[-,draw=black, thick] (eq) to (V);
\draw[->,draw=black,thick] (P) to (TR);
\draw[->,draw=black,thick] (V) to (AS);
\end{tikzpicture}
\caption{The goal of this paper is to mathematically formalize the above flowchart.}\label{fig:goal}
\end{figure}

The $\mathcal{N}=1$ super topological recursion can be used to compute (parts of the) correlation functions for a variety of examples related to 2d supergravity. For instance, we study applications to:
\begin{itemize}
\item $(2,4\ell)$-minimal superconformal models coupled to Liouville supergravity \cite{SL1,Beckers,SL2,SL3};
\item Super Jackiw–Teitelboim gravity \cite{SJT1,SJT2,SJT3,SW};
\item Supereigenvalue models in the Neveu-Schwarz sector \cite{SL1,Beckers,BO};
\item Supereigenvalue models in the Ramond sector \cite{C,O}.
\end{itemize}
For the first three examples in this list, it is known that the standard Eynard-Orantin topological recursion is sufficient to compute correlation functions, thanks to a non-trivial simplification first observed in \cite{Beckers}. However, as shown in \cite{O}, for the fourth example one needs the full $\mathcal{N}=1$ formalism.
Also, note that the first, third and fourth examples obey a truncation phenomenon \cite{Beckers,McArthur,O}, namely, correlation functions depend on fermions only up to quadratic order, which simplifies the super topological recursion. 

This paper is organized as follows. In Section~\ref{sec:SLE} we define local super spectral curves (Definition~\ref{def:SC}) and $\mathcal{N}=1$ abstract super loop equations (Definition~\ref{def:SLE}). 
In Section~\ref{sec:STR}, we solve the $\mathcal{N}=1$ abstract super loop equations through residue analysis, and construct a supersymmetric generalization of the Eynard-Orantin topological recursion, which we call the $\mathcal{N}=1$ super topological recursion (Proposition~\ref{prop:STR}). In Section~\ref{sec:SAS}, we transform the abstract super loop equations into differential constraints, and show that they form a super Airy structure, which comes with a unique partition function (Theorem~\ref{thm:main}). 
In Section~\ref{sec:Examples}, we discuss that (parts of the) correlation functions of the examples listed above can be computed by the $\mathcal{N}=1$ super topological recursion. Finally we conclude with a few open questions and future work. For the sake of brevity, the proofs of all theorems and propositions are given in Appendix~\ref{sec:Proofs}.

\subsubsection*{Acknowledgements}
We thank Rapha\"{e}l Belliard, Nitin Chidambaram, Thomas Creutzig, Naoki Genra, Shigenori Nakatsuka for various dicussions. We acknowledge the support of the Natural Sciences and Engineering Research Council of Canada. The work of KO is supported in part by the Engineering and Physical Sciences Research
Council under grant agreement ref. EP/S003657/2.

\section{Super Loop Equations}\label{sec:SLE}

In this section we fix notation, and introduce the notion of local super spectral curves. Given a local super spectral curve, we define $\mathcal{N}=1$ abstract super loop equations, which are the equations underlying the $\mathcal{N}=1$ super topological recursion. The presentation for the bosonic sector closely follows \cite{BEO} and \cite{HAS}.

\subsection{Local Spectral Curves}

We briefly review the notion of local spectral curves. Let us start with a symplectic vector space $V_{z}^B$ as
\begin{equation}
V_{z}^B:=\{\;\omega\in\mathbb{C}[z^{-1},z]]dz\;\;|\;\;\underset{z\rightarrow0}{\text{Res}}\,\omega(z)=0\;\},
\end{equation}
equipped with the following symplectic pairing $\Omega^B:V_z^B\times V_z^B\rightarrow\mathbb{C}$:
\begin{equation}
df_1,df_2\in V_z^B,\;\;\;\;\Omega^B(df_1,df_2)=\underset{z\rightarrow0}{\text{Res}} f_1(z)df_2(z).\label{inner product}
\end{equation}
We consider a Lagrangian subspace $V_{z}^{B+}=\mathbb{C}[[z]]dz\subset V_{z}^B$, and we choose a basis  $(d\xi_l)_{l>0}$ with
\begin{equation}
d\xi_l(z):=z^{l-1}dz,\;\;\;\;\;l\in\mathbb{Z}_{>0}.
\end{equation}

Given the Lagrangian subspace $V_z^{B+}$ with the choice of basis $(d\xi_l)_{l>0}$, we now choose another Lagrangian subspace $V_{z}^{B-}\subset V_{z}^B$ complementary to $V_z^{B+}$: we call this a choice of ``polarization''. That is, if we denote by $(d\xi_{-l})_{l>0}$ a basis of $V_z^{B-}$, then it satisfies:
\begin{equation}
\forall k,l\in\mathbb{Z}_{\neq0},\;\;\;\;\Omega^B(d\xi_k,d\xi_l)=\frac{\delta_{k+l,0}}{k}.
\end{equation}
Up to linear transformations, the above condition imposes that
\begin{equation}
d\xi_{-l}(z)=\frac{dz}{z^{l+1}}+\sum_{m>0}\frac{\phi_{lm}}{l}d\xi_m(z),\;\;\;\;l\in\mathbb{Z}_{>0},\label{bbasis}
\end{equation}
where we call the $\phi_{lm}=\phi_{ml}$ ``bosonic polarization parameters''. Note that the symmetry of $\phi_{lm}$ is required because of anti-symmetry of the symplectic pairing. 

Let us define a formal symmetric bidifferential $\omega_{0,2|0}$ in terms of the polarization as:
\begin{equation}
\omega_{0,2|0}(z_1,z_2|)=\frac{dz_1 dz_2}{(z_1-z_2)^2}+\sum_{k,l>0}\phi_{kl}\;d\xi_k(z_1) d\xi_l(z_2).
\end{equation}
Note that $\omega_{0,2|0}(z_1,z_2|)$ is not an element in $V_{z_1}^B\otimes V_{z_2}^B$ but rather
\begin{equation}
\omega_{0,2|0}(z_1,z_2|)-\frac{dz_1 dz_2}{(z_1-z_2)^2}\in V_{z_1}^{B+}\otimes V_{z_2}^{B+}.
\end{equation}
An important property of $\omega_{0,2|0}(z_1,z_2|)$ is that it works as a projection operator. That is, for any one-form $\omega\in V_{z}$ expanded as
\begin{equation}
\omega(z)=\sum_{l\neq0}c_l z^{l-1}dz,
\end{equation}
we get
\begin{equation}
\Omega^B(\omega_{0,2|0}(z,\cdot|),\omega)=\sum_{l>0}c_{-l}d\xi_{-l}(z)\in V_z^{B-}.\label{projection}
\end{equation}
In other words, it projects $\omega$ into $V_z^{B-}$. 
This can be easily checked by the fact that in the domain $|z_1|>|z_2|$, we can expand $\omega_{0,2|0}(z_1,z_2|)$ as
\begin{equation}
\omega_{0,2|0}(z_1,z_2|)=\sum_{l\geq1}ld\xi_{-l}(z_1)d\xi_l(z_2).\label{02z=0}
\end{equation}

The last ingredient in this section is an involution operator $\sigma:V^B\rightarrow V^B$ whose action is simply defined as
\begin{equation}
\sigma:z\mapsto-z.
\end{equation}
The basis of $V_z^{B+}$ is diagonal under $\sigma$, whereas the basis of $V_z^{B-}$ is generally not, due to nonzero polarization. 

With these ingredients, we can define a local spectral curve:

\begin{definition}[\cite{BEO,BS,HAS}]\label{def:C}
	A  \emph{local spectral curve with one component} consists of a symplectic vector space $V_z^B$, with a Lagrangian subspace $V_z^{B+}$, and the following data:\footnote{In \cite{HAS}, the more general case of automorphisms of arbitrary order was considered. Here we only consider involutions, which is consistent with the original topological recursion of Eynard and Orantin.}
\begin{itemize}
\item an involution operator $\sigma:V_z^B\rightarrow V_z^B$ whose action is defined as
\begin{equation}
\sigma:z\mapsto-z,
\end{equation}
\item a choice of ``dilaton shift parameters'' $(\tau_l)_{l>0}$, which can be encoded in a choice of a one-form $\omega_{0,1|0}\in V_z^{B+}$:\footnote{The condition $|\tau_1|+|\tau_3|>0$ is equivalent to what \cite{HAS} calls admissible.}
\begin{equation}
\omega_{0,1|0}(z)=\sum_{l>0}\tau_ld\xi_l(z),\;\;\;\;|\tau_1|+|\tau_3|>0,
	\end{equation}
\item a choice of bosonic polarization parameters, which can be encoded in a choice of symmetric bilinear differential $\omega_{0,2|0}$:
\begin{equation}
\omega_{0,2|0}(z_1,z_2|)=\frac{dz_1 dz_2}{(z_1-z_2)^2}+\sum_{k,l>0}\phi_{kl}\;d\xi_k(z_1) d\xi_l(z_2).
\end{equation}
\end{itemize}
\end{definition}

If one were to think of spectral curves in terms of branched coverings of Riemann surfaces, as in the original formulation of Eynard and Orantin, then the bosonic vector space $V_z^B$ would be interpreted as the space of differentials on an open neighbourhood of a simple ramification point of the branched covering, with $\omega_{0,2|0}$ being the Bergman kernel of the spectral curve, and $\sigma$ realizing the local involution that exchanges the two sheets of the branch cover near the ramification point. The name ``dilaton shift'' appears in the context of Airy structures \cite{HAS} rather than topological recursion, and we adapt it whenever we refer to a choice of parameters  $(\tau_l)_{l>0}$.

\subsection{Local Super Spectral Curves}

To define a local super spectral curve, we need two more ingredients: we need a vector space for fermions $V^F$, analogous to the bosonic vector space $V^B$, and a choice of fermionic polarization parameters encoded in a fermionic bilinear differential  $\omega_{0,0|2}$. 

We define a vector space $V^F(z,\theta)$ as:
\begin{equation}
	V^F_{z,\theta}:=\{\eta\in\mathbb{C}[z^{-1},z]]\;\Theta(z,\theta)\},
\end{equation}
where
\begin{equation}
	\Theta(z,\theta):=\left(\theta+zdz\frac{\partial}{\partial\theta}\right),
\end{equation}
and $\theta$ is a Grassman variable. We equip $V^F$ with a pairing $\Omega^F:V^F_{z,\theta}\times V^F_{z,\theta}\rightarrow\mathbb{C}$
\begin{equation}
\Omega^F(\eta_1,\eta_2):=\underset{z\rightarrow0}{\text{Res}}\;\eta_1(z,\theta)\eta_2(z,\theta),\label{F product}
\end{equation}
Note that $\Theta^2 = z dz$, hence, the residue makes sense.\footnote{We could have defined $\Theta = \sqrt{z dz}$ instead, without using the Grassman variable $\theta$, and the discussion below would still apply. Hence it appears that the Grassman variable may not be essential. It may however be important to define the notion of a \emph{global super spectral curve}. We hope to return to this point in the near future.}

\begin{remark}
We will often denote $\Theta(z,\theta)$ as $\Theta_z$ and $\Theta(z_i,\theta_i)$ as $\Theta_i$ for brevity. Also, we will often omit the $\theta$-dependence below, which should still be clear from the context.
\end{remark}


We extend the involution $\sigma$ in the definition of local spectral curves to:
\begin{equation}
\sigma:(z,\theta)\mapsto(-z,\theta).
\end{equation}
We note that $z dz$ is invariant under $\sigma$ so is $\Theta_z$.

Unlike the splitting of the vector space for bosons $V_z^B$ into two Lagrangian subspaces $V_z^{B+},V_z^{B-}$, we decompose $V_z^F$ into \emph{three} subspaces $V_z^{F+},V_z^{F0}$, and $V_z^{F-}$ as follows. Similar to $V_z^{B+}$, we define $V_z^{F+}=\{\eta\in \mathbb{C}[[z]]\,\Theta\}$, and we choose a basis $(\eta_l)_{l>0}$ with
\begin{equation}
\eta_l(z,\theta):=z^{l-1}\,\Theta,\;\;\;\;\;l\in\mathbb{Z}_{>0}.
\end{equation}
Next, we choose a polarization. First, we define $ V^{F\,0}$, which is a one-dimensional subspace whose basis $(\eta_0)$ is given by
\begin{equation}
\eta_0(z,\theta):=\left(\frac{1}{z}+\sum_{k>0}\psi_{0k}z^{k-1}\right)\,\Theta,\;\;\;\;\Omega^F(\eta_0,\eta_0)=1,\label{fbasis0}
\end{equation}
where $\psi_{0k}\in\mathbb{C}$. We call $\eta_0(z,\theta)$ the ``zero mode''. Finally, we let $V_z^{F-}$ be complementary to $V_z^{F+}\oplus V_z^{F0}$, with basis $(\eta_{-l})_{l\geq0}$ as
\begin{equation}
\eta_{-l}(z,\theta):=\left(\frac{1}{z^{l+1}}+\sum_{k\geq0}\psi_{lk}z^{k-1}\right)\,\Theta.\label{fbasis}
\end{equation}
We call the $\psi_{kl}$ the ``fermionic polarization parameters''. We require that
\begin{equation}
\forall k,l\in\mathbb{Z},\;\;\;\;\Omega^F(\eta_k,\eta_l)=\delta_{k+l,0}.
\end{equation}
This implies that
\begin{equation}
\psi_{00}=0,\;\;\;\;\psi_{kl}+\psi_{lk}+\psi_{0k}\psi_{0l}=0.
\end{equation}
That is, the $\psi_{kl}$ are not fully antisymmetric, due to the zero mode polarization, in contrast to the symmetry of the bosonic polarization parameters $\phi_{kl}$. 

We can encode the choice of polarization into a bilinear differential. We introduce an antisymmetric (fermionic) bilinear differential as
\begin{equation}
\omega_{0,0|2}(|z_1,z_2):=-\frac12\frac{z_1+z_2}{z_1-z_2}\frac{\Theta_1 \Theta_2}{z_1z_2}-\sum_{k,l\geq1}\frac{\psi_{k-1\;l-1}-\psi_{l-1\;k-1}}{1+\delta_{(k-1)(l-1),0}}\frac{\eta_l(z_1)  \eta_k(z_2)}{2z_1z_2}.
\end{equation}
Note that it is not an element of $V_{z_1}^F \otimes V_{z_2}^F$ but rather
\begin{equation}
z_1z_2\left(\omega_{0,0|2}(|z_1,z_2)+\frac12\frac{z_1+z_2}{z_1-z_2}\frac{\Theta_1 \Theta_2}{z_1z_2}\right)\in V_{z_1}^{F+}\otimes V_{z_2}^{F+}.
\end{equation}
In the domain $|z_1|<|z_2|$, it can be expanded as
\begin{equation}
\omega_{0,0|2}(|z_1,z_2)\rightarrow\sum_{l>0}\eta_{l}(z_1)\eta_{-l}(z_2)+\frac12\eta_0(z_1)\eta_0(z_2).
\end{equation}
It turns out that $\omega_{0,0|2}(|z_1,z_2)$ is a projection operator onto $V_z^{F0}\oplus V_z^{F-}$. That is, for any
\begin{equation}
\eta(z)=\sum_{l\in\mathbb{Z}}c_lz^{l-1}\,\Theta\in V_z^F,
\end{equation}
we have
\begin{equation}
\Omega^F(\omega_{0,0|2}(|\cdot,z)\,,\,\eta)=\sum_{l>0}c_l\eta_{-l}(z)+\frac12c_0\eta_0(z).\label{Fprojection}
\end{equation}

We are now ready to define a \emph{local super spectral curve}, which is a supersymmetric generalization of a local spectral curve (Definition~\ref{def:C}):

\begin{definition}\label{def:SC}
A  \emph{local super spectral curve $\mathcal{S_C}$ with one component} consists of a super symplectic vector space $V_z^B\oplus V_{z,\theta}^F$ with its maximum isotropic subspace $V_z^{B+}\oplus V_{z,\theta}^{F+}$ and the following data:
\begin{itemize}
\item an involution operator $\sigma:V_z^B\oplus V_{z,\theta}^F\rightarrow V_z^B\oplus V_{z,\theta}^F$ whose action is defined as
\begin{equation}
\sigma:(z,\theta)\mapsto(-z,\theta),
\end{equation}
\item a choice of dilaton shift, encoded in a choice of a one-form $\omega_{0,1|0}\in V_z^{B+}$
\begin{equation}
\omega_{0,1|0}(z)=\sum_{l>0}\tau_ld\xi_l(z),\;\;\;\;|\tau_1|+|\tau_3|>0,
\end{equation}
\item a choice of bosonic polarization, encoded in a symmetric bilinear differential $\omega_{0,2|0}$
\begin{equation}
	\omega_{0,2|0}(z_1,z_2|)=\frac{dz_1 dz_2}{(z_1-z_2)^2}+\sum_{k,l>0}\phi_{kl}\;d\xi_k(z_1) d\xi_l(z_2),\label{SSCw020}
\end{equation}
\item a choice of fermionic polarization, encoded in an anti-symmetric fermionic bilinear differential $\omega_{0,0|2}$
\begin{equation}
\omega_{0,0|2}(|z_1,z_2):=-\frac12\frac{z_1+z_2}{z_1-z_2}\frac{\Theta_1 \Theta_2}{z_1z_2}-\sum_{k,l\geq1}\frac{\psi_{k-1\;l-1}-\psi_{l-1\;k-1}}{1+\delta_{(k-1)(l-1),0}}\frac{\eta_l(z_1) \eta_k(z_2)}{2z_1z_2}.\label{SSCw002}
\end{equation}
\end{itemize}
\end{definition}
\begin{definition}
A local super spectral curve is said to be \emph{regular} if $\tau_1=0$, and \emph{irregular} if $\tau_1\neq0$.
\end{definition}
If one drops the vector space for fermions $V_{z,\theta}^F$ and the antisymmetric bilinear differential $\omega_{0,0|2}$ from the above definition, it reduces to Definition~\ref{def:C}.

\subsubsection{Local Super Spectral Curves With Several Components}
It is straightforward to generalize Definition~\ref{def:SC} to local super spectral curves with $c$ components by considering a vector space $\mathcal{V}_z^B\oplus\mathcal{V}_{z,\theta}^F=\mathbb{C}^c\otimes(V_z^B\oplus V_{z,\theta}^F)$ with $c\in\mathbb{Z}_{>0}$, similarly to \cite[Definition 5.7]{HAS}. That is, we associate a scalar product $\cdot$ to $\mathbb{C}^c$ with the standard orthogonal basis $(e_{\alpha})_{\alpha=1}^c$, and we define the symplectic products of $\mathcal{V}_z^B$ and $\mathcal{V}_{z,\theta}^F$ respectively as
\begin{equation}
\Omega^{B}(e_{\alpha_1}\otimes df_1,e_{\alpha_2}\otimes df_2):=\delta_{\alpha_1\alpha_2} \Omega^B(df_1,df_2),\;\;\;\Omega^{F}(e_{\alpha_1}\otimes \eta_1,e_{\alpha_2}\otimes \eta_2):=\delta_{\alpha_1\alpha_2} \Omega^F(\eta_1,\eta_2).
\end{equation}

We further define two subspaces $\mathcal{V}^{B+}=\mathbb{C}^c\otimes V^{B+}$ and $\mathcal{V}^{F+}=\mathbb{C}^c\otimes V^{F+}$, and choose their basis $(d\xi_{\alpha,l})$ and $(\eta_{\alpha,l})$ with $l\in\mathbb{Z}_{>0}$ and $\alpha\in\{1,\cdots,c\}$ as
\begin{equation}
d\xi_{\alpha,l}(z)=e_{\alpha}\otimes d\xi_l(z),\;\;\;\;\eta_{\alpha,l}(z)=e_{\alpha}\otimes \eta_l(z).
\end{equation}
Then, similar to the story with one component, we encode the information of polarizations of the remaining basis $(d\xi_{\alpha,l})$ and $(\eta_{\alpha,l})$ of $\mathcal{V}_z^B\oplus\mathcal{V}_{z,\theta}^F$ for $l\in\mathbb{Z}_{\leq0}$ in the definition of bilinear forms $\omega_{0,2|0}$ and  $\omega_{0,0|2}$. Thus, we have:

\begin{definition}\label{def:SCmulti}
A  \emph{local super spectral curve $\mathcal{S_C}$ with $c$ component} consists of a super symplectic vector space $\mathcal{V}_z^B\oplus \mathcal{V}_{z,\theta}^F$ with its maximum isotropic subspace $\mathcal{V}_z^{B+}\oplus \mathcal{V}_{z,\theta}^{F+}$ and the following data:
\begin{itemize}
\item a component-wise involution operator $\sigma_{\alpha}:\mathcal{V}_z^B\oplus \mathcal{V}_{z,\theta}^F\rightarrow \mathcal{V}_z^B\oplus \mathcal{V}_{z,\theta}^F$ whose action is defined for $l\in\mathbb{Z}$ and $\alpha,\beta\in\{1,\cdots,c\}$ by
\begin{equation}
\sigma_{\alpha}:d\xi_{\beta,l}(z)\mapsto d\xi_{\beta,l}((-1)^{\delta{\alpha\beta}}z),\;\;\;\;\eta_{\beta,l}(z)\mapsto \eta_{\beta,l}((-1)^{\delta{\alpha\beta}}z)
\end{equation}
\item a choice of dilaton shift, encoded in a choice of a one-form $\omega_{0,1|0}\in \mathcal{V}_z^{B+}$
\begin{equation}
\omega_{0,1|0}(z)=\sum_{\alpha=1}^c\sum_{l>0}\tau_{\alpha,l}d\xi_{\alpha,l}(z),\;\;\;\;\forall\alpha\;\;\;\;|\tau_{\alpha,1}|+|\tau_{\alpha,3}|>0,
\end{equation}
\item a choice of bosonic polarization, encoded in a symmetric bilinear differential $\omega_{0,2|0}$
\begin{equation}
	\omega_{0,2|0}(z_1,z_2|)=\sum_{\alpha=1}^c\frac{(e_{\alpha}\otimes dz_1)\otimes(e_{\alpha}\otimes dz_2)}{(z_1-z_2)^2}+\sum_{\alpha,\beta=1}^c\sum_{k,l>0}\phi_{kl}^{\alpha\beta}\;d\xi_{\alpha,k}(z_1) d\xi_{\beta,l}(z_2),\label{SSCw020multi}
\end{equation}
\item a choice of fermionic polarization, encoded in an anti-symmetric fermionic bilinear differential $\omega_{0,0|2}$
\begin{align}
\omega_{0,0|2}(|z_1,z_2):=&-\sum_{\alpha=1}^c\frac12\frac{z_1+z_2}{z_1-z_2}\frac{(e_{\alpha}\otimes \Theta_1)\otimes(e_{\alpha}\otimes \Theta_2)}{z_1z_2}\nonumber\\
&-\sum_{\alpha,\beta=1}^c\sum_{k,l\geq1}\frac{\psi_{k-1\;l-1}^{\alpha\beta}-\psi_{l-1\;k-1}^{\alpha\beta}}{1+\delta_{(k-1)(l-1),0}}\frac{\eta_{\alpha,l}(z_1) \eta_{\beta,k}(z_2)}{2z_1z_2}.\label{SSCw002multi}
\end{align}
\end{itemize}
\end{definition}

\begin{remark}
It is, however, not as straightforward to generalize the definition to higher order automorphisms (or spectral curves with higher order ramification): we leave this for future work.
\end{remark}

\subsection{$\mathcal{N}=1$ Abstract Super Loop Equations}

We now define $\mathcal{N}=1$ abstract super loop equations which we often call super loop equations for brevity. We again focus on local spectral curves with only one component.

Let us denote by $V_{z,\theta}^{F\,0,-}=V_{z,\theta}^{F\,0}\oplus V_{z,\theta}^{F-}$. Then for $g,n,m\in\mathbb{Z}_{\geq0}$ with $2g+n+2m>2$, we consider an infinite sequence of multilinear differentials $\omega_{g,n|2m}$ on a local super spectral curve $\mathcal{S_C}$ as
\begin{equation}
	\omega_{g,n|2m}\in \left(\bigotimes_{j=1}^nV_{z_j}^{B-}\right)\otimes \left(\bigotimes_{k=1}^{2m} V_{u_k,\theta_k}^{F\,0,-} \right).
\end{equation}
We impose that the $\omega_{g,n|2m}$ are symmetric under permutations of the first $n$ entries, and anti-symmetric under permutations of the last $2m$ entries. We assume no symmetry under permutations of some of the first $n$ entries with some of the last $2m$ entries.
Note that the $\omega_{g,n|2m}$ always have an even number of elements in $\bigotimes V_{u,\theta}^{F\,0,-}$. 

\begin{remark}
We say that the ``correlation functions'' $\omega_{g,n|2m}$ ``respect the polarization'', as they live in the subspaces $V_{z_j}^{B-}$ and $V_{u_k,\theta_k}^{F\, 0,-}$ defined by the choice of polarization in the local super spectral curve.
\end{remark}

Let us denote by $J,K$ a set of variables $J=(z_1,...)$ and $K=((u_1,\theta_1),..)$, and define the average of $\omega_{g,n|2m}$ under the involution $\sigma$ acting on each vector space as:
\begin{align}
\mathcal{L}_{g,n+1|2m}^B(z,J|K)=&\,\omega_{g,n+1|2m}(z,J|K)+\omega_{g,n+1|2m}(\sigma(z),J|K),\label{LB}\\
\mathcal{L}_{g,n|2m}^F(J|z,K)=&\,\omega_{g,n|2m}(J|z,K)+\omega_{g,n|2m}(J|\sigma(z),K),\label{LF}
\end{align}
where we dropped $\theta_i$'s from the arguments for brevity. Note that $|K|=2m$ in \eqref{LB} whereas $|K|=2m-1$ in \eqref{LF}. 

We further define the following quantities:
\begin{align}
\mathcal{Q}_{g,n|2m}^{FB}(J|z,K)=\;&\omega_{g-1,n+1|2m}(z,J|\sigma(z),K)+\omega_{g-1,n+1|2m}(\sigma(z),J|z,K)\nonumber\\
&+\sum_{g_1+g_2=g}\sum_{\substack{J_1\cup J_2=J \\ K_1\cup K_2=K}}(-1)^{\rho}\omega_{g_1,n_1+1|2m_1}(z,J_1|K_1)\,\omega_{g_2,n_2|2m_2}(J_2|\sigma(z),K_2)\nonumber\\
&+\sum_{g_1+g_2=g}\sum_{\substack{J_1\cup J_2=J \\ K_1\cup K_2=K}}(-1)^{\rho}\omega_{g_1,n_1+1|2m_1}(\sigma(z),J_1|K_1)\,\omega_{g_2,n_2|2m_2}(J_2|z,K_2),\label{QFB}
\end{align}
\begin{align}
\mathcal{Q}_{g,n+1|2m}^{BB}(z,J|K)=\;&\omega_{g-1,n+2|2m}(z,\sigma(z),J|K)\nonumber\\
&+\sum_{g_1+g_2=g}\sum_{\substack{J_1\cup J_2=J \\ K_1\cup K_2=K}}(-1)^{\rho}\omega_{g_1,n_1+1|2m_1}(z,J_1|K_1)\,\omega_{g_2,n_2+1|2m_2}(\sigma(z),J_2|K_2),\label{QBB}
\end{align}
\begin{align}
\mathcal{Q}_{g,n+1|2m}^{FF}(z,J|K)=\;&-\frac12\Bigl(\mathcal{D}_z\cdot\omega_{g-1,n|2m+2}(J|z,u,K)+\mathcal{D}_u\cdot\omega_{g-1,n|2m+2}(J|u,z,K)\Bigr)\Bigr|_{u=\sigma(z)}\nonumber\\
&+\frac12\sum_{g_1+g_2=g}\sum_{\substack{J_1\cup J_2=J \\ K_1\cup K_2=K}}(-1)^{\rho}\mathcal{D}_z\cdot\omega_{g_1,n_1|2m_1}(J_1|z,K_1)\, \omega_{g_2,n_2|2m_2}(J_2|\sigma(z),K_2)\nonumber\\
&+\frac12\sum_{g_1+g_2=g}\sum_{\substack{J_1\cup J_2=J \\ K_1\cup K_2=K}}(-1)^{\rho}\mathcal{D}_z\cdot\omega_{g_1,n_1|2m_1}(J_1|\sigma(z),K_1) \,\omega_{g_2,n_2|2m_2}(J_2|z,K_2),\label{QFF}
\end{align}
where for $\eta(z,\theta)=f(z)\Theta(z,\theta)\in V_{z,\theta}^F$, the derivative operator $\mathcal{D}_z$ is defined as
\begin{equation}
\mathcal{D}_z\cdot\eta(z,\theta)=df(z) \Theta(z,\theta)\in V_z^B\otimes V_{z,\theta}^F.
\end{equation}
Note that $(-1)^{\rho}=1$ if $K_1\cup K_2$ is an even permutation of $K$ and $(-1)^{\rho}=-1$ otherwise.

With these definitions, one can think of the abstract super loop equations on $\mathcal{S_C}$ as constrains imposing that the quantities above live in the ``plus'' subspaces of the vector spaces $V_z^B$ and $V_z^F$. More precisely: 


\begin{definition}\label{def:SLE}
Given a local super spectral curve $\mathcal{S_C}$, the \emph{$\mathcal{N}=1$ abstract super loop equations} are the following set of constraints:
\begin{enumerate}
\item \emph{linear bosonic loop equations}:
\begin{equation}
\mathcal{L}_{g,n+1|2m}^B(z,J|K)\in V_z^{B+},
\end{equation}
\item \emph{linear fermionic loop equations}:
\begin{equation}
\mathcal{L}_{g,n|2m}^F(J|z,K)\in V_z^{F+},
\end{equation}
\item \emph{quadratic bosonic loop equations}:
\begin{equation}
	\mathcal{Q}_{g,n+1|2m}^{BB}(z,J|K)+\mathcal{Q}_{g,n+1|2m}^{FF}(z,J|K)\in z V_z^{B+} \otimes z V_z^{B+},
\end{equation}
\item \emph{quadratic fermionic loop equations}:
\begin{equation}
\mathcal{Q}_{g,n|2m}^{FB}(J|z,K)\in z V_z^{B+}\otimes V_z^{F+}.
\end{equation}
\end{enumerate}
\end{definition}

These abstract super loop equations may seem rather \emph{ad hoc}. But they appear natural for a number of reasons. First, if one drops the fermionic vector space from consideration, the conditions (2) and (4) disappear, and conditions (1) and (3) reduce to the standard abstract loop equations. Second, the super loop equations that appear in the context of supereigenvalue models (see \cite{BO,O}) are particular cases of these abstract super loop equations. We will discuss this in Section~\ref{sec:Examples}. Third, and perhaps even more importantly, as we will see, these loop equations can be reformulated as differential constraints for a partition function $Z$, and these differential constraints take the form of a suitably polarized representation of the $\mathcal{N}=1$ super Virasoro algebra in the Neveu-Schwarz sector. In other words, the abstract super loop equations are a form of super Virasoro constraints. This is explored further in Section \ref{sec:SAS}.

\begin{remark}\label{rem:multi2}
For a local super spectral curves with $c$ component, recall from Definition~\ref{def:SCmulti} that the defining data carry an additional index $\alpha\in\{1,\cdots,c\}$. Accordingly, in this case we define \eqref{LB}-\eqref{QFF} and abstract super loop equations for each component $\alpha$ as in \cite[Definition 5.21]{HAS}. This makes sense because the involution $\sigma_{\alpha}$ is defined component-wise.
\end{remark}


\section{Super Topological Recursion}\label{sec:STR}
Our task is to prove that there exists a unique solution of super loop equations that respects the choice of polarization. If we assume existence, then it is relatively easy to construct a unique solution through residue analysis. This is what we do in this section. The resulting recursive formalism is a supersymmetric generalization of the Eynard-Orantin topological recursion, which we will call the \emph{$\mathcal{N}=1$ super topological recursion}.

As for existence of a solution, perhaps the simplest proof amounts to rewriting the abstract super loop equations as differential constraints, which take the form of a super Airy structure. We will discuss this approach in Section~\ref{sec:SAS}. Thus, for now, we assume existence of a solution to the abstract super loop equations.

Given a local super spectral curve $\mathcal{S_C}$, let us define what we call the recursion kernels:
\begin{align}
K^{BB}(z_0,z,\sigma(z))=\,&\frac{\int^{z}_{0}\omega_{0,2|0}(z_0,\cdot|)}{\omega_{0,1|0}(z|)-\omega_{0,1|0}(\sigma(z)|)},\\
K^{BF}(z_0,z,\sigma(z))=\,&\frac{\omega_{0,0|2}(|z,z_0)-\frac12\eta_0(z)\eta_0(z_0)}{\omega_{0,1|0}(z|)-\omega_{0,1|0}(\sigma(z)|)}.\label{BFK}
\end{align}
We note that for each local super spectral curve, those kernels are uniquely defined. The first one is the standard recursion kernel in the Eynard-Orantin topological recursion, whereas the second one is new and incorporates fermions. 

In the limit $z\rightarrow0$, the numerator of \eqref{BFK} becomes
\begin{equation}
\sum_{k\geq1}\eta_{k}(z)\eta_{-k}(z_0),
\end{equation}
hence this factor works as a projection to $V^{F\,-}_{z_0}$. Remark that this is different from the bilinear differential $\omega_{0,0|2}(|z_0,z)$ which projects onto $V^{F\, 0,-}_{z_0}$ as discussed in  \eqref{Fprojection}. Indeed, one needs to be very careful with the fermionic zero modes (this was also noticed in the construction of super Airy structures in \cite{SAS}). 
As we will show in Appendix~\ref{sec:proof:STR}, it turns out that this projection is exactly what we need to solve the abstract super loop equations through residue analysis, and develop a supersymmetric generalization of the Eynard-Orantin topological recursion.

\begin{proposition}\label{prop:STR}
Let  $\tilde{\mathcal{Q}}_{g,n+1|2m}^{BB,FF,BF}$ denote respectively all the terms on the right hand side of \eqref{QFB}-\eqref{QFF} \emph{except} the terms involving $\omega_{0,1|0}$. If there exists a solution to the  $\mathcal{N}=1$ abstract super loop equations that respects the polarization, then it is uniquely constructed recursively by the following formulae:
\begin{align}
\omega_{g,n+1|2m}(z_0,J|K)=\;&\underset{z\rightarrow0}{{\rm Res}}\,K^{BB}(z_0,z,\sigma(z))\left(\tilde{\mathcal{Q}}_{g,n+1|2m}^{BB}(z,J|K)+\tilde{\mathcal{Q}}_{g,n+1|2m}^{FF}(z,J|K)\right),\label{BR}\\
\omega_{g,n|2m+2}(J|u_1,u_2,K)=\;&\hat{\omega}_{g,n|2m+2}(J|u_1,u_2,K)-\eta_0(u_1)\underset{z\rightarrow0}{{\rm Res}}\,\hat{\omega}_{g,n|2m+2}(J|u_2,z,K)\eta_0(z),\label{FR}
\end{align}
where
\begin{equation}
\hat{\omega}_{g,n|2m+2}(J|u_1,u_2,K)=\underset{z\rightarrow0}{{\rm Res}}\,K^{FB}(u_1,z,\sigma(z))\tilde{\mathcal{Q}}_{g,n|2m+2}^{FB}(J|z,u_2,K).\label{FR1}
\end{equation}
\end{proposition}

See Appendix~\ref{sec:proof:STR} for the proof.

\begin{remark}
	\eqref{BR} and \eqref{FR} do not guarantee \emph{a priori} that the $\omega_{g,n|2m}$ are symmetric under permutations of the first $n$ entries, and anti-symmetric under permutations of the last $2m$ entries. 
Also, for $n m\neq0$, one can compute $\omega_{g,n|2m}$ from either \eqref{BR} or \eqref{FR}, and it is not clear \emph{a priori} that they coincide. In other words, the solution constructed as above may not even exist. Existence of solution is proven in the next section in terms of super Airy structures.
\end{remark}

\begin{remark}\label{rem:multi3}
For a local spectral curve with $c$ components, recall from Remark ~\ref{rem:multi2} that the abstract super loop equations are labelled by an additional index $\alpha\in\{1,\cdots,c\}$ due to component-wise involutions $\sigma_{\alpha}$. As a result, the recursion kernels are defined for each component $\alpha$, and one should take summation over $\alpha$ from $1$ to $c$ in order to obtain the correct differentials $\omega_{g,n|2m}$. This is similar to \cite[Definition 5.19]{HAS}.
\end{remark}

\section{Super Airy Structures}\label{sec:SAS}

In this section we show how one can solve the abstract super loop equations through the framework of super Airy strucctures \cite{SAS}. The idea is to rewrite the super loop equations as differential constraints on a partition function, and show that these constraints satisfy the properties of a super Airy structure, which guarantees existence and uniqueness of the partition function.
Let us start by briefly reviewing the notion of super Airy structures. See \cite{SAS} for more details.

\subsection{Review of Super Airy Structures}

Let $U=U_0\oplus U_1\oplus\mathbb{C}^{0|1}$ be a super vector space of dimension $d+1$ over $\mathbb{C}$ (the super vector space could be infinite-dimensional, but for simplicity of presentation we will assume here that it has finite dimension). We define $\{x^i\}_{i\in I}$ to be linear coordinates on $U_0\oplus U_1$ where $I=\{1,...,d\}$ with $x^0$ to be the coordinate of the extra $\mathbb{C}^{0|1}$, and their parity is defined such that $|x^i|=0$ if $x^i\in U_0$, $|x^i|=1$ if $x^i\in U_1$, and $|x^0|=1$. Note that $x^0\in\mathbb{C}^{0|1}$ plays an analogous role to $\eta_0(z)\in V_z^{F\,0}$ that appeared in Section~\ref{sec:STR}. Furthermore, let us denote by
\begin{equation}
\mathcal{D}_{\hbar}(U)=\mathbb{C}[[\hbar,x^0,\hbar\partial_{x^0},\{x^i\}_{i\in I},\{\hbar\partial_{x^i}\}_{i\in I}]]
\end{equation}
the completed algebra of differential operators acting on $U$, and we introduce a $\mathbb{Z}$-grading by
\begin{equation}
\deg(x^0)=\deg(x^i)=1,\;\;\;\deg(\hbar\partial_{x^0})=\deg(\hbar\partial_{x^i})=1,\;\;\;\;\deg(\hbar)=2.
\end{equation}

\begin{definition}[{\cite[Definition 2.3]{SAS}}]\label{def:SAS}
A \emph{super Airy structure} is a set of differential operators $\{H_i\}_{i\in I}\in\mathcal{D}_{\hbar}(U)$ such that:
\begin{enumerate}
\item for each $i\in I$, $H_i$ is of the form
\begin{equation} 
H_i=\hbar\partial_{x^i}-P_i,\label{form}
\end{equation}
where $P_i\in\mathcal{D}_{\hbar}(U)$ has degree greater than 1 with $|P_i|=|x^i|$,
\item there exists $f_{ij}^k\in\mathcal{D}_{\hbar}(U)$ such that
\begin{equation}
[H_i,H_j]=\hbar\sum_{k\in I} f_{ij}^k \,H_k,\label{left}
\end{equation}
where $[\cdot,\cdot]$ is a super-commutator.
\end{enumerate}
\end{definition}

It is crucial that the $x^0$-dependence appears only in the $\{P_i\}_{i\in I}$, but not in the degree 1 term (there is no $H_0$). We call $x^0$ the \emph{extra variable}. Accordingly, the dimension of the super vector space $U$ is one more than the number of $\{H_i\}_{i\in I}$. We note that there is no notion of extra variables in the standard, nonsupersymmetric, formalism of Airy structures.

\begin{theorem}[{\cite[Theorem 2.10]{SAS}}]\label{thm:SAS1}
Given a super Airy structure  $\{H_i\}_{i\in I}$, there exists a unique formal power series $\hbar F(x)\in\mathbb{C}[[\hbar,x^0,(x^i)_{i\in I}]]$ (up to addition of terms in $\mathbb{C}[[\hbar]]$) such that:
\begin{enumerate}
\item $\hbar F(x)$ has no term of degree 2 or less,
\item every term in $\hbar F(x)$ has even parity,
\item it satisfies $H_i\,e^{F}=0$.
\end{enumerate}
\end{theorem}

$Z := e^F$ is called the \emph{partition function} and $F$ \emph{the free energy}. Note that $e^F$ is not a power series in $\hbar$, and so one should replace condition (3) by $e^{-F}\,H_i\,e^{F}=0$, which gives a power series in $\hbar$. However, as is standard, we write $H_i\,e^{F}=0$ for brevity. 

Explicitly, $F$ can be expanded as follows
\begin{equation}
F=\sum_{g,n\geq0}^{2g+n>2}\frac{\hbar^{g-1}}{n!}\sum_{i_1,...,i_n\in \{0,I\}}F_{g,n}(i_1,...,i_n)\prod_{k=1}^nx^{i_k},
\end{equation}
where the restriction on the sum that $2g + n >2$ comes from the first condition in Theorem~\ref{thm:SAS1}. 
$F_{g,n}(i_1,...,i_n)$ is $\mathbb{Z}_2$-symmetric under permutations of indices.

\subsection{Super Loop Equations and Super Airy Structures}

Our goal is now to turn the abstract super loop equations into differential constraints for a partition function $Z$. More precisely, we expand the correlation functions $\omega_{g,n|2m}$ satisfying the abstract super loop equations in the basis defined previously as:
\begin{equation}
\omega_{g,n|2m}(J|K)=\sum_{\substack{i_1,...,i_n>1\\j_1,...,j_{2m}\geq0}}F_{g,n|2m}(i_1,...,i_n|j_1,...,j_{2m})\bigotimes_{k=1}^n d\xi_{-i_k}(z_k)\otimes\bigotimes_{l=1}^{2m}\eta_{-j_l}(u_l,\theta_l)
\end{equation}
Then, we want to show that the $F_{g,n|2m}$ that appear in this decomposition are the coefficients of the free energy $F$ for some super Airy structure. If we can show that, by Theorem \ref{thm:SAS1}, it will ensure existence and uniqueness of the free energy, and hence of the solution of the abstract super loop equations given by the super topological recursion in Proposition \ref{prop:STR}.

To construct the relevant super Airy structures, we proceed as follows. We take both the bosonic and fermionic vector spaces $U_0,U_1$ to be countably infinite dimensional. Also, we explicitly distinguish bosonic and fermionic coordinates, namely, we denote by $\{x^1,x^2,...\}$ and $\{\theta^1,\theta^2,...\}$ the coordinates on $U_0$ and $U_1$ respectively and $\theta^0\in\mathbb{C}^{0|1}$ is treated as the extra variable. In particular, all $\{\theta^0,\theta^1,\theta^2,...\}$ are Grassmann variables. 

We then define $\{J_a\}_{a\in\mathbb{Z}}$ and $\{\Gamma_a\}_{a\in\mathbb{Z}}$ by:
\begin{align}
\forall a\in\mathbb{Z}_{>0},&\;\;\;\;J_{a}=\hbar\frac{\partial}{\partial x^a},\;\;\;\;J_0=0,\;\;\;\;J_{-a}=ax^a,\\
\forall a\in\mathbb{Z}_{>0},&\;\;\;\;\Gamma_{a}=\hbar\frac{\partial}{\partial\theta^{a}},\;\;\;\;\Gamma_0=\frac{\theta^0}{2}+\hbar\frac{\partial}{\partial\theta^0},\;\;\;\;\Gamma_{-a}=\theta^{a},
\end{align}
It is easy to see that the $J_a$ are a basis for the Heisenberg algebra, while the $\Gamma_a$ are a basis for the Clifford algebra:
\begin{equation}
[J_a,J_b]=a\,\hbar\,\delta_{a+b,0},\;\;\;\;[J_a,\Gamma_{b}]=0,\;\;\;\;\{\Gamma_{a},\Gamma_{b}\}=\hbar\,\delta_{a+b,0}.
\end{equation}

Using those, we define the following set of differential operators (where $: \cdots :$ denotes normal ordering):
\begin{align}
n\in\mathbb{Z}_{\geq-1},\;\;\;\;L_{2n}&=\frac12\sum_{j\in\mathbb{Z}}(-1)^{j-1} : J_{-j}J_{2n+j} : + \frac{\hbar}{4}\delta_{n,0}\nonumber\\
&\;\;\;\;+ \frac12\sum_{j\in\mathbb{Z}}(-1)^{j}(n+j) : \Gamma_{-j}\Gamma_{j+2n} :,\label{L1}\\
m\in\mathbb{Z}_{\geq-1},\;\;\;\;G_{2m+1}&=\sum_{j\in\mathbb{Z}} (-1)^{j-1} :J_{-j}\Gamma_{j+2m+1} :.\label{G1}
\end{align}
It is easy to show that for $n,m\in\mathbb{Z}_{\geq-1}$ and $i\in\mathbb{Z}_{\geq1}$, these operators satisfy the following commutation relations:
\begin{equation}
[L_{2n},J_{2i}]=2 i \hbar J_{2n+2i},\;\;\;\;[G_{2m+1},J_{2i}]=2i \hbar \Gamma_{2i+2m+1},\label{comm1}
\end{equation}
\begin{equation}
[L_{2n},\Gamma_{2i-1}]=(n+2i-1) \hbar \Gamma_{2n+2i-1},\;\;\;\;\{G_{2m+1},\Gamma_{2i-1}\}=\hbar J_{2i+2m+2},
\end{equation}
\begin{align}
[L_{2n},L_{2m}]=&2\hbar(n-m)\biggl(L_{2n+2m}+\sum_{j\in\mathbb{Z}}: J_{-2j}J_{2n+2m+2j} :\nonumber\\
&+\sum_{j\in\mathbb{Z}}(n+m+2j+1) : \Gamma_{-2j-1}\Gamma_{2j+2n+2m+1} :\biggr),\\
	[L_{2n},G_{2m+1}]=&\hbar(n-2m-1)\biggl(G_{2n+2m+1}+2\sum_{j\in\mathbb{Z}}: J_{-2j}\Gamma_{2n+2m+2j+1} :\biggr),\\
\{G_{2n+1},G_{2m+1}\}=&2\hbar\biggl(L_{2n+2m+2}+\sum_{j\in\mathbb{Z}}: J_{-2j}J_{2n+2m+2j+2} :\nonumber\\
&+\sum_{j\in\mathbb{Z}}(n+m+2j+2) : \Gamma_{-2j-1}\Gamma_{2j+2n+2m+3} :\biggr).\label{comm5}
\end{align}
This is a natural extension of the $\mathcal{N}=1$ super Virasoro algebra in the Neveu-Schwarz sector by the first-order differential operators $J_{2i}$ and $\Gamma_{2i-1}$.

We now introduce the notion of a dilaton shift and polarization in the context of super Airy structures. For $\tau_l,\phi_{kl},\psi_{kl}\in\mathbb{C}$, we consider a differential operator $\Phi$ as:
\begin{equation}
\Phi:=\exp\left(\frac{1}{\hbar}\left(\sum_{l>0}\frac{\tau_l}{l}J_l+\sum_{l,k>0}\frac{\phi_{kl}}{2kl}J_kJ_l+\sum_{k,l\geq0}\frac{\psi_{kl}}{2}\Gamma_k\Gamma_l\right)\right),\label{Phi}
\end{equation}
We then define dilaton-shifted and polarized operators $\{\tilde L_{2n},\tilde G_{2m+1} \}$ for $n,m \in \mathbb{Z}_{\geq -1}$:
\begin{equation}
	\tilde L_{2n} = \Phi L_{2n} \Phi^{-1}, \qquad \tilde G_{2m+1} = \Phi G_{2m+1} \Phi^{-1}.
\end{equation}
Note that conjugating by $\Phi$ simply acts by shifting the modes $J_{-i}$ and $\Gamma_{-i}$ as:
\begin{equation}
	J_{-i} \mapsto J_{-i}+\tau_i+\sum_{k\geq1}\frac{\phi_{ik}}{k}J_k,\;\;\;\; \Gamma_{-i} \mapsto \Gamma_{-i}+\sum_{k\geq0}\psi_{ki}\Gamma_k,\label{Phi on modes}
\end{equation}
where we conventionally defined $\tau_i=\phi_{ik}=\psi_{i-1,k}=0$ for $i\in\mathbb{Z}_{\leq0}$. Notice that one can find the differential operators $\{\tilde L_{2n},\tilde G_{2m+1} \}$ from the defining data of a super spectral curve and vise versa. This is the natural generalization of the dilaton shift considered in \cite{HAS}.


With this under our belt, we get the following result:

\begin{proposition}\label{prop:SASpolar}
For $i\in\mathbb{Z}_{>0}$, consider the set $\mathcal{S_A}$ of differential operators
\begin{equation}
\mathcal{S_A}=\{H^1_i,F^1_i,H^2_i,F^2_i\},\label{S_A}
\end{equation}
where
\begin{equation}
	H^1_i=J_{2i},\;\;\;\;F^1_i=\Gamma_{2i-1},\;\;\;\;H^2_i=\tilde L_{2i-\epsilon-1},\;\;\;\;F^2_i=\tilde G_{2i-\epsilon}.\label{HJHJ}
\end{equation}
We set $\epsilon = 1$ if $\tau_1 = 0$, and $\epsilon = 3$ otherwise. Then the differential operators in $\mathcal{S_A}$ form a super Airy structure, with $\theta_0$ the extra variable.
\end{proposition}
See Appendix~\ref{sec:Proof1} for the proof.


Since the differential operators in $\mathcal{S_A}$ form a super Airy structure, Theorem~\ref{thm:SAS1} implies that there exists a unique partition function $Z$ and free energy $F = \log Z$ in the form:
\begin{equation}
F=\sum_{g,n,m\geq0}^{2g+n+2m>2}\frac{\hbar^{g-1}}{n!(2m)!}\sum_{\substack{i_1,...,i_n>1\\j_1,...,j_{2m}\geq0}}F_{g,n|2m}(i_1,...,i_n|j_1,...,j_{2m})\prod_{k=1}^nx^{i_k}\prod_{l=1}^{2m}\theta^{j_l},\label{SASF}
\end{equation}
and such that
\begin{equation}
\forall \;i\in\mathbb{Z}_{>0},\;\;\;\;H^1_iZ=F^1_iZ=H^2_iZ=F^2_iZ=0.
\end{equation}
Note that $F_{g,n|2m}$ is symmetric under permutations of the $n$ first entries, anti-symmetric under permutations of the last $2m$ entries, with no further symmetry.



Our goal now is to relate this super Airy structure to the abstract super loop equations. This is the essence of the following theorem.

\begin{theorem}\label{thm:main}
	~
	\begin{enumerate}
		\item
			Consider the super Airy structure $\mathcal{S_A}$ in Proposition \ref{prop:SASpolar}, defined in terms of the dilaton shift and polarization parameters $\tau_l, \phi_{kl}$ and $\psi_{kl}$. Let
			\begin{equation}
				F_{g,n|2m}(i_1,\ldots , i_n | j_1, \ldots, j_{2m} )
				\end{equation}
				be the coefficients of the unique free energy $F$ associated with this super Airy structure $\mathcal{S_A}$. 
\item
	Let $\mathcal{S}_C$ be a super spectral curve defined in terms of the same dilaton shift and polarization parameters $\tau_l, \phi_{kl}$ and $\psi_{kl}$. Consider an infinite sequence of multilinear differentials $\omega_{g,n|2m}$ that respect the polarization:
	\begin{equation}
		\omega_{g,n|2m}\in \left(\bigotimes_{j=1}^nV_{z_j}^{B-}\right)\otimes \left(\bigotimes_{k=1}^{2m} V_{u_k,\theta_k}^{F\,0,-} \right),
\end{equation}
and that satisfy the abstract super loop equations Definition \ref{def:SLE}. We expand the differentials in terms of the basis in the definition of super spectral curves as: 
\begin{equation}
	\omega_{g,n|2m}(J|K)=\sum_{\substack{i_1,...,i_n>1\\j_1,...,j_{2m}\geq0}}\hat F_{g,n|2m}(i_1,...,i_n|j_1,...,j_{2m})\bigotimes_{k=1}^n d\xi_{-i_k}(z_k)\otimes\bigotimes_{l=1}^{2m}\eta_{-j_l}(u_l,\theta_l) .\label{thm}
\end{equation}
\end{enumerate}
Then, for all $g,n,m$, and indices $i_1, \ldots, i_n$ and $j_1, \ldots, j_{2m}$,
\begin{equation}
	\hat F_{g,n|2m}(i_1,...,i_n|j_1,...,j_{2m}) = 	F_{g,n|2m}(i_1,\ldots , i_n | j_1, \ldots, j_{2m} ).
\end{equation}

\end{theorem}

We give the proof in Appendix~\ref{sec:Proof2}. Concretely, what we are doing is reformulating the abstract super loop equations as differential constraints satisfied by the partition function $Z$, which take the form of the super Airy structure $\mathcal{S_A}$ defined in terms of the polarization of the super spectral curve.


An immediate corollary of this theorem is that a solution to the abstract super loop equations that respects the polarization exists. As a result, it must be given by the super topological recursion in Proposition \ref{prop:STR}.
\begin{corollary}\label{coro:main}
	There exists a solution to the abstract super loop equations that respects the polarization, and it is uniquely constructed by the $\mathcal{N}=1$ super topological recursion of Proposition \ref{prop:STR}.
\end{corollary}

\begin{remark}\label{rem:multi4}
For a local spectral curves with $c$ component, all one has to do is to prepare $c$ copies of super Virasoro operators. Then, it is straightforward to generalise Proposition~\ref{prop:SASpolar} and Theorem~\ref{thm:main} with several components.
\end{remark}

\subsection{Going Back to the Super Loop Equations}\label{sec:going back}

We end this section with an important remark. In the construction of the super Airy structure $\mathcal{S_A}$, the operators $H^1_i$ and $F^1_i$ for $i \in \mathbb{Z}_{>0}$ are just derivatives:
\begin{equation}
	H^1_i = J_{2i} = \hbar \frac{\partial}{\partial x^{2i}}, \qquad F^1_i = \Gamma_{2i-1} = \hbar \frac{\partial}{\partial \theta^{2i-1}}.
\end{equation}
Thus, the differential constraints $H^1_i Z = F^1_i Z = 0$ impose that the partition function $Z$ does not depend on the variables $x^{2i}$ and $\theta^{2i-1}$ for all $i \in \mathbb{Z}_{>0}$. As a consequence, we can reduce the differential operators (this is similar to the reduction considered in Section 2.2.3 of \cite{HAS}) by setting $J_{2i} =0$ and $\Gamma_{2i-1} = 0$ for all $i \in \mathbb{Z}$. The resulting differential operators (after rescaling) form a representation of the $\mathcal{N}=1$ super Virasoro algebra in the Neveu-Schwarz sector.

In particular, if we choose a trivial polarization, by setting 
$\tau_k=\phi_{kl}=\psi_{kl}=0$ except for $\tau_3=1$, the operators precisely agree with the representation given in Section 4.2.6 of \cite{SAS}.

This was in fact part of the motivation for introducing the particular abstract super loop equations that we considered in Definition \ref{def:SLE}. On the one hand, we wanted our abstract super loop equations to be natural generalizations of the standard bosonic ones, and to include as particular cases the super loop equations of supereigenvalue models. But, on the other hand, we were also looking for super loop equations that correspond to the (suitably polarized) differential constraints associated with the super Airy structures realized as representations of the super Virasoro algebras considered in \cite{SAS}. Those motivations resulted in Definition \ref{def:SLE}.

\section{Examples}\label{sec:Examples}
In this section, we will apply the $\mathcal{N}=1$ super topological recursion (equivalently super Airy structures) to compute (parts of) correlation functions of the examples listed below:
\begin{itemize}
\item $(2,4\ell)$-minimal superconformal models coupled to Liouville supergravity,
\item Super Jackiw–Teitelboim gravity,
\item Supereigenvalue Models in the Neveu-Schwarz sector,
\item Supereigenvalue models in the Ramond sector .
\end{itemize}
We will approach the first two examples with the techniques of super Airy structures. Concretely, we will show an interesting relation to ordinary Airy structures as an extension of \cite{Beckers}, which helps us with describing the first two examples in terms of super Airy structures with suitable dilaton shift and polarization. In contrast, the last two examples are described in the framework of the $\mathcal{N}=1$ super topological recursion. That is, we show that their correlation functions satisfy the abstract super loop equations on a certain local super spectral curve, hence they are uniquely constructed thanks to Proposition~\ref{prop:STR}.

\subsection{Relation between Airy Structures and Super Airy Structures}\label{sec:reduction}

We investigate a relation between Airy structures and super Airy structures with vanishing polarization but with arbitrary choice of dilaton shift. This naturally leads us to the first two examples.

To do so, let us first define a set of operators $\check L_{2n}$ by
\begin{equation}
n\in\mathbb{Z}_{\geq-1},\;\;\;\;\check L_{2n}=\frac12\sum_{j\in\mathbb{Z}}(-1)^{j-1} : J_{-j}J_{2n+j} : + \frac{\hbar}{8}\delta_{n,0}.
\end{equation}
 $\check L_{2n}$ are same as the first line of \eqref{L1} except the last term which is now $\hbar/8$ instead of $\hbar/4$. We then construct dilaton-shifted operators $\check L_{2n}^{\tau}$ by taking conjugate as
 \begin{equation}
\check L_{2n}^{\tau}=\exp\left(\frac{1}{\hbar}\sum_{l>0}\frac{\tau_l}{l}J_l\right)\,\check L_{2n}\,\exp\left(-\frac{1}{\hbar}\sum_{l>0}\frac{\tau_l}{l}J_l\right).
\end{equation}
We further define $\check H^2_i=\check L^{\tau}_{2i-\epsilon-1}$. Recall the definition of $H^1_i$ from \eqref{HJHJ}, then it is shown in \cite{HAS} that a set $\mathcal{A}^{\tau}=\{H_i^1,\check H^2_i\}_{i\in\mathbb{Z}_{>0}}$ of differential operators forms an Airy structure with one component, and as a consequence, there is a unique partition function annihilated by those differential operators. (See \cite{HAS} for the definition of Airy structure in general. Alternatively, it is sufficient for our purpose if one just drops all Grassmann variables in Definition~\ref{def:SAS} from consideration.)

Let us now consider another set $\mathcal{S_A}^{\tau}$ of differential operators given in \eqref{S_A} with the choice of dilaton shift parameters being exactly the same $\tau_l$ in $\mathcal{A}^{\tau}$ and polarization being trivial, $\phi_{kl}=\psi_{kl}=0$.  We also choose $\epsilon$ in $\mathcal{S_A}^{\tau}$ to be the same as that in $\mathcal{A}^{\tau}$. Then,  Proposition~\ref{prop:SASpolar} immediately implies that  $\mathcal{S_A}^{\tau}$ forms a super Airy structure. Somewhat surprisingly, we find the following relation between the Airy structure $\mathcal{A}^{\tau}$ and the super Airy structure $\mathcal{S_A}^{\tau}$:
\begin{proposition}\label{prop:reduction}
Let $F(\mathcal{A}^{\tau})$ and $F(\mathcal{S_A}^{\tau})$ be the free energy associated with the Airy structure $\mathcal{A}^{\tau}$, and that with the super Airy structure $\mathcal{S_A}^{\tau}$ defined above respectively. Then, order by order in $\hbar$, we have
\begin{equation}
F_g(\mathcal{S_A}^{\tau})=2^g\left(F_g(\mathcal{A}^{\tau})-\frac12\sum_{i,j\geq0}\theta^{2i}\theta^{2j}\frac{\partial^2F_g(\mathcal{A}^{\tau})}{\partial x^{2i+1}\partial x^{2j-1}}\right)+\mathcal{O}(\theta^4),\label{reduction}
\end{equation}
where $\mathcal{O}(\theta^4)$ and higher terms vanish if $\epsilon=3$.
\end{proposition}
The proof is given in Section~\ref{sec:proof of reduction} in detail, but let us give a few remarks about this proposition. This type of relation is first observed in \cite{Beckers} for the case with $\tau_l=\delta_{l,3}$ in line with supereigenvalue models, and \cite{McArthur} proved that $\mathcal{O}(\theta^4)$ or higher terms in $\theta$ vanish. That is, the free energy truncates at quadratic order with respect to Grassmann variables\footnote{It remains to be seen whether truncation holds for $\epsilon=1$ too.}. However, since the formula in \cite{Beckers} was not written in the form of \eqref{reduction}, this point was not realized in \cite{SAS} in relation to super Airy structures. Proposition~\ref{prop:reduction} is an extension of \cite{Beckers} to arbitrary dilaton shift including irregular ones. It remains to be investigated how general we can extend this type of relation with nonzero polarization.

With Proposition~\ref{prop:reduction} in our hands, we are able to discuss the first two examples in the list above.

\subsubsection{$(2,4\ell)$-minimal superconformal models coupled to Liouville supergravity}

It was shown \cite{SL1,Beckers,SL2,SL3} that the continuum limit of  supereigenvalue models in the Neveu-Schwarz sector (cases without the continuum limit will be presented shortly) describe $(2,4\ell)$-minimal superconformal models coupled to Liouville supergravity, which turns out to be a solution of a supersymmetric extension of the KdV-hierarchy too \cite{SKdV}. 
After an appropriate transformation, the free energy of a corresponding super Airy structure  becomes the generating function of correlation functions of $(2,4\ell)$-minimal superconformal models coupled to Liouville supergravity. See \cite{SL1,SL3} for more details about the necessary transformation.

\begin{proposition}\label{prop:SKdV}
Let $F^{{\rm LSG}}(\mathcal{S_A}^{\tau_3})$ be the free energy associated with the super Airy structure with $\tau_l=\delta_{l,3}$, $\phi_{kl}=\psi_{kl}=0$, and $\epsilon=3$. Then, after an appropriate transformation, $F^{{\rm LSG}}(\mathcal{S_A}^{\tau_3})$  becomes the generating function of correlation functions of $(2,4\ell)$-minimal superconformal models coupled to Liouville supergravity.
\end{proposition}
This is an old story discussed in \cite{SL1,Beckers,SL2,SL3}. Since their presentation is different from the style of this paper, we give a short justification in Appendix~\ref{sec:just:SKdV}.

\subsubsection{ Super Jackiw–Teitelboim gravity}
Thanks to Proposition~\ref{prop:reduction}, the  free energy $F(\mathcal{A}^{\tau})$ encodes the same information as the free energy $F^{(0)}(\mathcal{S_A}^{\tau})$ where the superscript $(k)$ denotes the order of Grassmann variables. This includes the Kontsevich-Witten $\tau$-function \cite{K,W}, the Brezin-Gross-Witten $\tau$-funtion \cite{BG,GW}, and Mirzakhani's recursion for volumes of moduli spaces of Riemann surfaces \cite{Mirzakhani1,Mirzakhani2}. An interesting question arises: is there any super Airy structure $\mathcal{S_A}^{\tau}$ such that not only $F^{(0)}(\mathcal{S_A}^{\tau})$  but also $F^{(2)}(\mathcal{S_A}^{\tau})$ have enumerative interpretation? Even though we do not have any promised example, let us discuss a possible candidate.

Recently, Stanford and Witten investigated super Jackiw–Teitelboim gravity \cite{SJT1,SJT2,SJT3} and showed in \cite{SW} that volumes of moduli spaces of super Riemann surfaces can be computed by utilizing the Eynard-Orantin topological recursion. They derived that the spectral curve has no polarization, and dilaton shift parameters $(\tau_l)_{l>0}$ are encoded in the following one-form
\begin{equation}
\omega_{0,1|0}(z|)=\sqrt{2}\cos(2\pi z)dz=\sum_{l>0}\tau_ld\xi_l(z).
\end{equation}
If we apply Proposition~\ref{prop:reduction} with the dilaton shift given above, we know the role of $F_{g,n|0}$ thanks to \cite{SW}. How about $F_{g,n|2}$? Can we find their enumerative meanings in terms of moduli spaces of super Riemann surfaces, or physical meanings in terms of super Jackiw–Teitelboim gravity? It remains to be investigated, but the study of Ramond punctures might be a relevant starting point.

\subsection{Supereigenvalue Models}

Supereigenvalue models (see \cite{SL1,Beckers,C1,C2,C,BO,O} and references therein) are supersymmetric generalizations of Hermitian matrix models. It is known that (parts of) correlation functions of these models satisfy super loop equations, and their recursive system has been verified in \cite {BO} for the Neveu-Scwarz sector and in \cite{O} for the Ramond sector. However, their super loop equations appear to look differently from each other so do the resulting recursive formulae. A benefit of our abstract formalism is that regardless the sector, correlation functions satisfy the abstract super loop equations, and we can apply the super topological recursion to construct a unique solution. Therefore, the super topological recursion is a unifying recursive formalism --  one can treat both the Neveu-Schwarz and Ramond sector in the same footing.



\subsubsection{Neveu-Schwarz Sector}
The local super spectral curve for the Neveu-Schwarz sector consists of \emph{two} components. Since a global expression is known thanks to \cite{BO}, it is the sufficient if we present how to determine all the dilaton shift parameters $\tau_{\alpha,l}$ and polarization parameters $\phi_{k,l}^{\alpha,\beta}, \psi_{k-1,l-1}^{\alpha,\beta}$ with $k,l\in\mathbb{Z}_{>0}$ and $\alpha,\beta\in\{+,-\}$ from the spectral curve given in \cite{BO}.

Let us first define polynomials $x_{\pm}\in\mathbb{C}[z]$ and formal power series $u_{\pm}\in\mathbb{C}[[z]]$ encoded in the following form:
\begin{equation}
x_{\pm}(z)=\pm1+\frac{z^2}{2},\;\;\;\;u_{\pm}(z)=\pm1+\frac{z^2}{2}+z\sqrt{\pm1+\frac{z^2}{4}},
\end{equation}
where the equality for $u_{\alpha}(z)$ should be undertstood as a Taylor expansion at $z = 0$. The sign of the square root in $u_{\alpha}(z)$ is not an issue here because that exactly amounts to the action of the involution in the definition of local super spectral curves. Note that $u_{\alpha}(z)$ comes from the global coordinate of a hyperelliptic curve given in \cite{BO} whereas $z$ can be thought of a local coordinate in the neighbourhood of one of the ramification points. Then for any polynomial $M(x)$ with $M(\pm1)\neq0$, $\tau_{\alpha,l}$ are determined by the following term-by-term equation in $z$
\begin{equation}
\sum_{l>0}\tau_{\alpha,l}z^{l-1}dz=\frac12 M(x_{\alpha}(z)) \left(u_{\alpha}(z)-\frac{1}{u_{\alpha}(z)}\right)zdz,\label{tauNS}
\end{equation}
where one should expand the right hand side at $z=0$.

Next, let us define a bilinear differential $B(u_1,u_2)$ as
\begin{equation}
B(u_1,u_2)=\frac{du_1 du_2}{(u_1-u_2)^2}.
\end{equation}
Then, for $\alpha,\beta\in\{+,-\}$, bosonic polarization parameters $\phi_{k,l}^{\alpha,\beta}$ are determined by the following term-by-term equations in $z_1,z_2$:
\begin{equation}
\frac{dz_1 dz_2}{(z_1-z_2)^2}\delta_{\alpha,\beta}+\sum_{k,l>0}\phi_{k,l}^{\alpha,\beta}z_1^{k-1}z_2^{l-1}dz_1dz_2=B(u_{\alpha}(z_1),u_{\beta}(z_2)).\label{phiNS}
\end{equation}
Similarly, fermionic polarization parameters are determined by the following term-by-term equations in $z_1,z_2$:
\begin{align}
-\frac12\frac{z_1+z_2}{z_1-z_2}\frac{\delta_{\alpha\beta}}{z_1z_2}-\sum_{k,l\geq1}\frac{\psi_{k-1\;l-1}^{\alpha,\beta}-\psi_{l-1\;k-1}^{\alpha,\beta}}{1+\delta_{(k-1)(l-1),0}}z_1^{l-1}z_2^{k-1}=\frac{x_{\alpha}(z_1)-x_{\beta}(z_2)}{dx_{\alpha}(z_1)dx_{\beta}(z_2)}B(u_{\alpha}(z_1),u_{\beta}(z_2)),\label{psiNS}
\end{align}
where the right hand sides of \eqref{phiNS} and \eqref{psiNS} should be expanded at $z_1, z_2=0$. Note that for $\alpha\neq\beta$, one can indeed show that the right hand sides of \eqref{phiNS} and \eqref{psiNS} are regular at $z_1 = z_2$, which is consistent with Definition~\ref{def:SCmulti}.

\begin{proposition}[Neveu-Schwarz sector]\label{prop:NS}

Let us consider a local super spectral curve $\mathcal{S_C}^{NS}$ with two components whose dilaton shift and polarization parameters are given by \eqref{tauNS}, \eqref{phiNS}, and \eqref{psiNS}.
Then, for $2g+n+2m>2$, $\omega_{g,n|2m}$ constructed from the $\mathcal{N}=1$ super topological recursion on $\mathcal{S_C}^{NS}$ correspond to (fermionic-coupling independent) correlation functions of supereigenvalue models in the Neveu-Schwarz sector.
\end{proposition}
See Appendix~\ref{sec:NS} for the proof. Note that an analogous formula to \eqref{reduction} is known to hold for supereigenvalue models in the Neveu-Schwarz sector, and it was a key fact for \cite{BO} to present a recursive formula. On the other hand, the $\mathcal{N}=1$ super topological recursion gives a unique solution without referring to such a simplification.



\subsubsection{Ramond Sector}

The local spectral curve for the Ramond sector consists of only one component due to a (somewhat surprising) supersymemtric cancellation observed in \cite{O}. Since the global expression is known thanks to \cite{O}, it is again sufficient to present how to determine the defining parameters of the corresponding super spectral curve.

Let us first define a polynomial $x\in\mathbb{C}[z]$ and a formal power series $u\in\mathbb{C}[[z]]$ encoded in the following form:
\begin{equation}
x(z)=1+\frac{z^2}{2},\;\;\;\;u(z)=z(2+z^2)^{-\frac12}.
\end{equation}
where the equality for $u(z)$ should be undertstood as a Taylor expansion at $z = 0$. Then, similar to the Neveu-Schwarz sector, the dilaton shift and bosonic polarization parameters are determined by the following term-by-term equations:
\begin{equation}
\sum_{l>0}\tau_{l}z^{l-1}dz= M(x(z))u(z)zdz,\label{tauR}
\end{equation}
\begin{equation}
\frac{dz_1 dz_2}{(z_1-z_2)^2}+\sum_{k,l>0}\phi_{k,l}z_1^{k-1}z_2^{l-1}dz_1dz_2=B(u(z_1),u(z_2)),\label{phiR}
\end{equation}
where the right hand sides should be expanded at $z,z_1, z_2=0$. On the other hand, the fermionic polarization parameters are determined by the following term-by-term equation in $z_1,z_2$:
\begin{align}
&-\frac12\frac{z_1+z_2}{z_1-z_2}\frac{1}{z_1z_2}-\sum_{k,l\geq1}\frac{\psi_{k-1\;l-1}-\psi_{l-1\;k-1}}{1+\delta_{(k-1)(l-1),0}}z_1^{l-1}z_2^{k-1}\nonumber\\
&\hspace{40mm}=-\frac{(u(z_1)+u(z_2))(1-u(z_1)u(z_2))}{4u(z_1)u(z_2)(u(z_1)-u(z_2))\sqrt{x(z_1)}\sqrt{x(z_2)}},\label{psiR}
\end{align}
Note that $\sqrt{x(z)}$ does not create any issue regarding branch cuts because \eqref{psiR} is a valid equation only in the neighourbood of $z=0$ $(x=1)$. This is another advantage of considering a local super spectral curve -- one of difficulties in the Ramond sector is the appearance of square roots, and \cite{O} had to consider a variant of correlation functions in order to evaluate them as single-valued differentials on a Riemann surface.

\begin{proposition}[Ramond sector]\label{prop:R}
Let us consider a local super spectral curve $\mathcal{S_C}^{R}$ with one components whose dilaton shift and polarization parameters are given by \eqref{tauR}, \eqref{phiR}, and \eqref{psiR}.
Then, for $2g+n+2m>2$, $\omega_{g,n|2m}$ constructed from the $\mathcal{N}=1$ super topological recursion on $\mathcal{S_C}^{R}$ correspond to (fermionic-coupling independent) correlation functions of supereigenvalue models in the Ramond sector.
\end{proposition}
The proof is given in Appendix~\ref{sec:R}.

\begin{remark}
The current formalism is not sufficient to incorporate fermionic couplings in supereigevalue models. We suspect that investigating fermionic couplings helps with developing the notion of ``global super spectral curves''. We are hoping to return to it in the near future. 
\end{remark}

\subsection{Comments on Truncation}

It is proven for minimal superconformal models and supereigenvalue models in both sectors that their correlation functions (equiv. free energy) truncate at quadratic order in fermionic variables   -- and the authors suspect that this applies to all local super spectral curves with vanishing polarizations. However, this does not hold for a more general class of local super spectral curves. In fact, if we consider a local super spectral curve with nonzero polarization given as
\begin{align}
\omega_{0,1|0}(z|)&=z^2dz,\\
\omega_{0,2|0}(z_1,z_2|)&=\left(\frac{1}{(z_1-z_2)^2}+\phi_{11}\right)dz_1 dz_2,\\
\omega_{0,0|2}(|z_1,z_2)&=-\frac12\frac{z_1+z_2}{z_1-z_2}\frac{\Theta_1\Theta_2}{z_1z_2},
\end{align}
then we find that
\begin{equation}
\omega_{2,0|4}(|z_1,z_2,z_3,z_4)=\phi_{11}^3\sum_{i_l=0}^3\epsilon_{i_1i_2i_3i_4}\eta_{-2i_1}(z_1)\eta_{-2i_2}(z_2)\eta_{-2i_3}(z_3)\eta_{-2i_4}(z_4),
\end{equation}
where $\epsilon_{i_1i_2i_3i_4}$ is completely antisymmetric under the permutation of the indices and it is normalized as $\epsilon_{0123}=1$. A general analysis on truncation phenomena remains to be investigated and seems rather complex. However, it is easy to show that for $m\in\mathbb{Z}_{\geq2}$, $\omega_{0,0|2m}$ and $\omega_{0,1|2m}$ vanish for any local super spectral curve: 
\begin{proposition}\label{prop:truncation}
\begin{equation}
\forall\;m\in\mathbb{Z}_{\geq2},\;\;\;\;\omega_{0,0|2m}=\omega_{0,1|2m}=0.
\end{equation}
\end{proposition}
See Appendix~\ref{sec:truncation} for the proof.

\section{Conclusion and Future Work}\label{sec:Conclusion}
We have formalized the flowchart in Figure~\ref{fig:goal}  through Definition~\ref{def:SLE}, Proposition~\ref{prop:STR}, and Theorem~\ref{thm:main}. There is a one-to-one correspondence between $\omega_{g,n|2m}$ on a local super spectral curve $\mathcal{S_C}$ and $F_{g,n|2m}$  associated with a super Airy structure $\mathcal{S_A}$. We have then discussed that four examples related to 2d supergravity fit into this new framework, and we are seeking more. Let us conclude  with listing open questions and future work.

\subsection*{Global super spectral curves}
\cite{BO,O} showed that the full recursion of supereigenvalue models in both the Neveu-Schwarz and Ramond sector requires one more initial datum; a Grassmann-valued polynomial equation. 
These observations suggest a possibility of defining a \emph{global} super spectral curve which comes with Grassmann-valued parameters. Note that every global spectral curve can be described by a local spectral curves with multiple components by looking at an open neighbourhood of every ramification point. If we believe that this holds in  a supersymmetric realm, then how can we consider a local super spectral curve compatible with possible Grassmann-valued parameters? Since the current formalism is based on $\mathbb{C}$-valued vector spaces $V^B,V^F$, a fundamental extension seems necessary to construct a formalism equipped with Grassmann parameters.

Even though we do not have any rigorous idea, let us mention a few expectations. First, supereigenvalue models suggest to introduce $\omega_{0,0|1}$ -- the Grassmann-valued counterpart of $\omega_{0,1|0}$ --, and as a result, there would possibly be ``fermionic dilaton shift'' as well as nonzero $\omega_{g,n|2m+1}$. On the other hand, from a super Airy structure point of view, we would have to allow $F_{g,n|m}$ to be Grassmann-valued in such a way that the partition function $Z$ is still bosonic. Thus, in particular, we need to generalize super Airy structures defined in \cite{SAS}. It remains to be investigated how to make sense of these insights with technical details.

\subsection*{Higher generalization}
\cite{HAS} has shown a correspondence between $\mathcal{W}$-algebra and the Bouchard-Eynard topological recursion which involves higher orders of ramification. A natural question is whether we can upgrade their work with supersymmetry. In terms of the super topological recursion, this would potentially mean that we generalize the involution $\sigma:z\mapsto-z$ to an automorphism $\sigma:z\mapsto e^{2\pi i/q}z$ for some $q\in\mathbb{Z}_{\geq2}$. The super $\mathcal{W}$-algebra counterpart, however, is not so clear how we should generalize, and we are hoping to return to this point in the near future. This is indeed under investigation joint with N. Chidambaram, T. Creutzig, N. Genra, and S. Nakatsuka. While we were finishing up this paper, a new paper \cite{Chen:2020hml} appeared on the arXiv that discusses a $\mathcal{W}$-algebra and supereigenvalue models in the Ramond sector. It is interesting to see how our formalism relates to theirs.

\subsection*{Enumerative geometry}
Following the work of Stanford-Witten \cite{SW}, Norbury very recently developed in \cite{N} an intersection theory associated with moduli spaces of super Riemann surfaces. Even though the recursion in this story is the standard recursion of  Eynard and Orantin, it is interesting to see whether the $\mathcal{N}=1$ super topological recursion plays an additional role, in particular, whether $\omega_{g,n|2}$ admit some enumerative interpretation. One good starting point would be the study of Ramond punctures. Furthermore, the analysis in \cite{SW,N} reduces down to computations in ``reduced spaces'' of moduli spaces of super Riemann surfaces \cite{W5}. Such reduced spaces can be obtained by setting all odd moduli to zero, and they turn out to be moduli spaces of ordinary Riemann surfaces with the extra data of spin structures. Importantly, the bosonic part of the Teichm\"{u}ller space does not see the extra spin structures, hence they are the same as usual Teichm\"{u}ller space \cite{N}. It is interesting to investigate whether this fact relates to Proposition~\ref{prop:reduction}. Putting another way, intersection theory on more general moduli spaces may require a recursive formalism beyond the Eynard-Orantin topological recursion, and the $\mathcal{N}=1$ super topological recursion may play a key role.

\newpage
\appendix

\section{Proofs}\label{sec:Proofs}

\subsection{Proof of Proposition~\ref{prop:STR}}\label{sec:proof:STR}

The proof of \eqref{BR} closely follows how \cite{BEO,BS} prove the standard  local topological recursion. Given a local super spectral curve, let us assume existence of solutions of the abstract super loop equations that respects the polarization. Since $K^{BB}(z_0,z,\sigma(z))$ has at most a simple pole at $z=0$, the quadratic bosonic loop equations imply that
\begin{equation}
\forall n, m\in\mathbb{Z}_{\geq0},\;\;\;\;\underset{z\rightarrow0}{{\rm Res}}\,K^{BB}(z_0,z,\sigma(z))\biggl(\mathcal{Q}_{g,n+1|2m}^{BB}(z,J|K)+\mathcal{Q}_{g,n+1|2m}^{FF}(z,J|K)\biggr)=0.
\end{equation}
Let us focus on terms involving $\omega_{0,1|0}$ on the left hand side. They appear in the form:
\begin{align}
&\underset{z\rightarrow0}{{\rm Res}}\,K^{BB}(z_0,z,\sigma(z))\biggl(\omega_{0,1|0}(z|)\omega_{g,n+1|2m}(\sigma(z),J|K)+\omega_{0,1|0}(\sigma(z)|)\omega_{g,n+1|2m}(z,J|K)\biggr)\nonumber\\
&=\underset{z\rightarrow0}{{\rm Res}}\,K^{BB}(z_0,z,\sigma(z))\biggl(-\left(\omega_{0,1|0}(z|)-\omega_{0,1|0}(\sigma(z)|)\right)\omega_{g,n+1|2m}(z,J|K)\nonumber\\
&\hspace{50mm}+\omega_{0,1|0}(z|)\mathcal{L}_{g,n+1|2m}^B(z,J|K)\biggr)\label{p1},
\end{align}
where we used \eqref{LB} in the equality. The linear bosonic loop equations guarantee that the last term in \eqref{p1} does not contribute to the residue at $z=0$. Furthermore, $\omega_{0,1|0}(z|)-\omega_{0,1|0}(\sigma(z)|)$ on the right hand side cancels the denominator in the recursion kernel. In summary, we get
\begin{equation}
\eqref{p1}=-\underset{z\rightarrow0}{{\rm Res}}\int^{z}_{0}\omega_{0,2|0}(z_0,\cdot|)\,\omega_{g,n+1|2m}(z,J|K)=-\omega_{g,n+1|2m}(z_0,J|K),
\end{equation}
where we used \eqref{projection} in the second equality. This proves \eqref{BR}.

Similarly, the quadratic fermionic loop equations imply
\begin{equation}
\forall n, m\in\mathbb{Z}_{\geq0},\;\;\;\;\underset{z\rightarrow0}{{\rm Res}}\,K^{FB}(u_1,z,\sigma(z))\mathcal{Q}_{g,n|2m+2}^{FB}(J|z,u_2,K)=0.\label{p2}
\end{equation}
Note that the right hand side vanishes thanks to the $-\frac12\eta_0(z)\eta_0(u_1)$ in the recursion kernel \eqref{BFK}. 
One can repeat the same procedure as we did in \eqref{p1}, with the help of the linear fermionic loop equations instead, then terms involving $\omega_{0,1|0}$ on the left hand side of  \eqref{p2} become
\begin{align}
-\underset{z\rightarrow0}{{\rm Res}}\left(\omega_{0,0|2}(|z,u_1)-\frac12\eta_0(z)\eta_0(u_1)\right)\omega_{g,n|2m+2}(J|z,u_2,K)=-\hat{\omega}_{g,n|2m+2}(J|u_1,u_2,,K).
\end{align}
This gives \eqref{FR1}. Remark that due to the $-\frac12\eta_0(z)\eta_0(u_1)$ factor in the recursion kernel, $\hat{\omega}_{g,n|2m+2}(J|u_1,u_2,K)$ recovers $\omega_{g,n|2m+2}(J|u_1,u_2,K)$ \emph{except} terms that depend on $\eta_0(u_1)$. Since fermionic entries are antisymmetric under their permutations, however, one can indeed supplement this missing $\eta_0(u_1)$-dependence precisely by the second term in \eqref{FR}. It is clear that \eqref{BR} together with \eqref{FR} are recursive for $\omega_{g,n|2m}$ in $2g+n+2m$, hence we have constructed all $\omega_{g,n|2m}$ starting with a local super spectral curve, subject to the assumption of existence of solution. This completes the proof.

\subsection{Proof of Proposition~\ref{prop:SASpolar}}\label{sec:Proof1}
Since conjugation by $\Phi$ does not change the commutation relations, it is clear from the commutation relations \eqref{comm1}-\eqref{comm5} that the differential operators in $\mathcal{S_A}$ satisfy property (2) in Definition~\ref{def:SAS}. Thus it suffices if we show that (linear combinations of) the differential operators in $\mathcal{S_A}$ satisfies property (1) in Definition~\ref{def:SAS}. We present a proof for $\epsilon=3$ ($\tau_1=0$), and we normalize $\tau_3=1$ -- the following discussions can be straightforwardly applied to the case with $\epsilon=1$ ($\tau_1\neq0$).

It is shown in \eqref{Phi on modes} that $H_i^1,F_i^1$ remain unchanged under conjugation by $\Phi$, hence, they automatically satisfy property (1). On the other hand, $H_i^2,F_i^2$ are expanded in terms of $(J_i,\Gamma_i)_{i\in\mathbb{Z}}$ as follows:
\begin{align}
H_i^2&=\sum_{k\in\mathbb{Z}_{\geq2}}(-1)^{k-1} \tau_k J_{2i+k-4}+\frac12\sum_{j,k\in\mathbb{Z}}\left(C_i^{j,k|}:J_jJ_k:+C_i^{|j,k}:\Gamma_j\Gamma_k:\right)+\hbar D_i,\\
F_i^2&=\sum_{k\in\mathbb{Z}_{\geq2}}(-1)^{k-1} \tau_k \Gamma_{2i+k-3}+\sum_{j,k\in\mathbb{Z}}C_i^{j|k}:J_j\Gamma_k:,
\end{align}
where
\begin{align}
C_i^{j,k|}&=(-1)^{j-1}\delta_{j+k,2i-4}+(-1)^{j-1}\frac{\phi_{j-2i+4,k}}{k}+(-1)^{k-1}\frac{\phi_{j,k-2i+4}}{j}+\delta_{i,1}\frac{\phi_{1,j}\phi_{1,k}}{jk},\label{C1}\\
C_i^{|j,k}&=(-1)^j\frac{k-j}{2}\delta_{j+k,2i-4}+\delta_{i,1}(\psi_{2,j}\psi_{0,k}-\psi_{0,j}\psi_{2,k})\nonumber\\
&\;\;\;\;+(-1)^j(k-i+2)\psi_{j,k-2i+4}-(-1)^k(j-i+2)\psi_{k,j-2i+4},\label{C2}\\
C_i^{j|k}&=(-1)^k\delta_{j+k,2i-3}+(-1)^k\frac{\phi_{j,k-2i+3}}{j}-(-1)^j\psi_{k,j-2i+3}+\delta_{i,1}\frac{\phi_{j,1}}{j}\psi_{k,0},\label{C3}\\
D_i&=\frac14\delta_{i,2}+\left(\frac12\phi_{1,1}+\frac12\psi_{0,2}\right)\delta_{i,1},\label{D1}
\end{align}
where we conventionally defined $\phi_{i,k}=\psi_{i,k}=0$ for $i\in\mathbb{Z}_{<0}$. We then introduce a parity-preserving linear transformation to $\hat{H}^2_i,\hat{F}^2_i$ as
\begin{align}
\forall\,i\in\mathbb{Z}_{\geq1},\;\;\;\;&\hat{H}^2_i=H_i^2+\sum_{k\geq1}\tau_{2k} H^1_{i+k-2}-\sum_{k\geq2}\tau_{2k+1} H^2_{i+k-1},\label{linear2}\\
&\hat{F}^2_i=F_i^2+\sum_{k\geq1}\tau_{2k} F^1_{i+k-2}-\sum_{k\geq2}\tau_{2k+1} F^2_{i+k-1},\label{linear3}
\end{align}
where we conventionally defined $H_0^1=J_0=0$. In particular, the degree 1 term in each $\hat{H}^2_i,\hat{F}^2_i$ reads, respectively,
\begin{equation}
\hat{H}^2_i=\hbar \frac{\partial}{\partial x^{2i-1}}+(\deg2 \text{ terms}),\;\;\;\;\hat{F}^2_i=\hbar \frac{\partial}{\partial \theta^{2i}}+(\deg2 \text{ terms}).
\end{equation}
Therefore, the set $\mathcal{\hat{S}_A}=\{H_i^1,F_i^1,\hat{H}_i^2,\hat{F}_i^2\}$ is a super Airy structure. Note that $\theta^0$ does not appear in the degree 1 term, hence it plays the role of the extra variable.

\subsection{Proof of Theorem~\ref{thm:main}}\label{sec:Proof2}
The differential operators in $\mathcal{\hat{S}_A}$ above and $\mathcal{S_A}$ are related by the  linear transformations \eqref{linear2} and \eqref{linear3}. As a consequence, if $Z$ is the associated unique partition function of the super Airy structure $\mathcal{\hat{S}_A}$, then it is also annihilated by all the operators in $\mathcal{S_A}$. In other words, property (2) is required only up to linear transformations (independent of the choice of basis). Thus, we can equivalently consider differential constrains derived from the differential operators in $\mathcal{S_A}$ instead of those in $\mathcal{\hat{S}_A}$. It turns out that this is easier in practice to show a relation to abstract super loop equations.

With this remark in mind, the proof consists of two parts:
\begin{itemize}
\item[\textit{Part 1}] given a super Airy structure $\mathcal{S_A}$, obtain a sequence of equations that the associated free energy $F(\mathcal{S_A})$ satisfies,
\item[\textit{Part 2}] starting with the corresponding local super spectral curve  $\mathcal{S_C}$, expand the abstract super loop equations that respect the polarization in the $(\xi_{k},\eta_{l})$-basis and show that the expanding coefficients satisfy exactly the same sequence of equations.
\end{itemize}

\subsubsection{Part 1}\label{sec:step1}

Let us denote by $Z,F$ the unique partition function and free energy associated with a super Airy structure $\mathcal{\hat{S}_A}$ Then, $Z$ is annihilated by all differential operators in $\mathcal{S_A}$. It is easy to show that $H_i^iZ=0$, $F_i^1Z=0$ give respectively
\begin{equation}
F_{g,n+1|2m}(2i,J|K)=0,\;\;\;\;F_{g,n|2m+2}(J|2i-1,K)=0,\label{linear Z}
\end{equation}
for any $g,n,m\in\mathbb{Z}_{\geq0}$ (recall that $F_{0,1|0}=F_{0,2|0}=F_{0,0|2}=0$ by definition). Here we denote by $J=\{i_1,i_2,...\}$ a collections of positive integers and by $K=\{j_1,j_2,...\}$ by a collection of nonnegative integers. We will come back later that these equations are in agreement with the linear bosonic and fermionic loop equations.

We now consider the equations derived from $H_i^2Z=F_i^2Z=0$. In order to express all terms in a compact way, let us first define the following notations:
\begin{align}
\Xi_{g,n+1|2m}^{(0)}[i,J|K]&=\sum_{p\geq1}(-1)^{p-1}\tau_pF_{g,n+1|2l}(2i+p-4,J|K),\\
\Xi_{g,n|2m}^{(0)}[J|i,K]&=\sum_{p\geq1}(-1)^{p-1}\tau_pF_{g,n+1|2m}(J|2i+p-3,K),\\
\Xi_{g,n|2m}^{(2)}[k,l,J|K]&=F_{g-1,n+2|2m}(k,l,J|K)\nonumber\\
&\;\;\;\;+\sum_{g_1+g_2=g}\sum_{\substack{J_1\cup J_2=J \\ K_1\cup K_2=K}}(-1)^{\rho}F_{g_1,n_1+1|2m_1}(k,J_1|K_1)F_{g_2,n_2+1|2m_2}(l,J_2|K_2),\\
\Xi_{g,n|2m}^{(2)}[J|k,l,K]&=-F_{g-1,n|2m+2}(J|k,l,K)\nonumber\\
&\;\;\;\;+\sum_{g_1+g_2=g}\sum_{\substack{J_1\cup J_2=J \\ K_1\cup K_2=K}}(-1)^{\rho}F_{g_1,n_1|2m_1}(J_1|k,K_1)F_{g_2,n_2|2m_2}(J_2|l,K_2),
\end{align}
\begin{align}
\Xi_{g,n|2m}^{(2)}[k,J|l,K]&=F_{g-1,n+1|2m}(k,J|l,K)\nonumber\\
&\;\;\;\;+\sum_{g_1+g_2=g}\sum_{\substack{J_1\cup J_2=J \\ K_1\cup K_2=K}}(-1)^{\rho}F_{g_1,n_1+1|2m_1}(k,J_1|K_1)F_{g_2,n_2|2m_2}(J_2|l,K_2).
\end{align}
Then order by order in $\hbar$ as well as in variables $x^j,\theta^j$, we find from $H_i^2Z=F_i^2Z=0$ a sequence of constraints on the free energy $F$. For $\Xi_{g,n|2m}^{(0)}$ with $2g+n+2m=3$, we have
\begin{align}
\forall\;j,k\in\mathbb{Z}_{\geq1},\;\;\;\;0=&\Xi_{0,3|0}^{(0)}[i,j,k|]+jkC_i^{-j,-k|},\label{chi=31}\\
\forall\;j,k\in\mathbb{Z}_{\geq0},\;\;\;\;0=&\Xi_{0,1|2}^{(0)}[i|j,k]+\frac{1}{1+\delta_{j,0}+\delta_{k,0}}C_i^{|-j,-k},\label{chi=32}\\
\forall\;j\in\mathbb{Z}_{\geq1},\;\;\;\;k\in\mathbb{Z}_{\geq0},\;\;\;\;0=&\Xi_{0,1|2}^{(0)}[j|i,k]+\frac{j}{1+\delta_{k,0}}C_i^{-j|-k},\label{chi=33}\\
0=&\,\Xi^{(0)}_{1,1|0}[i|]+D_i.\label{chi=34}
\end{align}
Note that from the definitions \eqref{C1}, \eqref{C2}, and \eqref{C3}, one easily finds
\begin{equation}
C_i^{-j,-k|}=\delta_{i,1}\delta_{j,1}\delta_{k,1},\;\;\;\;C_i^{|-j,-k}=\delta_{i,1}(\delta_{j,2}\delta_{k,0}-\delta_{j,0}\delta_{k,2}),\;\;\;\;C_i^{-j|-k}=\delta_{i,1}\delta_{j,1}\delta_{k,0},\label{chi=35}
\end{equation}
for \eqref{chi=31}, \eqref{chi=32}, and \eqref{chi=33} respectively. For $\Xi_{g,n+1|2m}^{(0)}$ with $2g+n+2m\geq4$, we get from $H_i^2Z=0$ that
\begin{align}
0=&\,\Xi_{g,n+1|2m}^{(0)}[i,J|K]+\sum_{k,l\geq0}\left(C_i^{k,l|}\Xi_{g,n|2m}^{(2)}[k,l,J|K]+C_i^{|k,l}\Xi_{g,n|2m}^{(2)}[J|k,l,K]\right)\nonumber\\
&+\sum_{k\geq0}\left(\sum_{l=1}^ni_lC_i^{-i_l,k|}F_{g,n|2m}(k,J\backslash i_l|K)+\sum_{l=1}^{2m}(-1)^{l-1}\frac{C_i^{|-j_l,k}}{{1+\delta_{j_l,0}}}F_{g,n|2m}(J|k,K\backslash j_l)\right).\label{BSAS}
\end{align}
And for $\Xi_{g,n|2m}^{(0)}$ with $2g+n+2m\geq4$, we find from $F_i^2Z=0$ that
\begin{align}
0=&\,\Xi_{g,n|2m}^{(0)}[J|i,K]+\sum_{k,l\geq0}C_i^{k|l}\Xi_{g,n|2m}^{(2)}[k,J|l,K]\nonumber\\
&+\sum_{k\geq0}\left(\sum_{l=1}^ni_lC_i^{-i_l|k}F_{g,n-1|2m}(J\backslash i_l|k,K)+\sum_{l=1}^{2m-1}(-1)^{l-1}\frac{C_i^{k|-j_l}}{1+\delta_{j_l,0}}F_{g,n+1|2m-2}(k,J|K\backslash j_l)\right).\label{FSAS}
\end{align}
See Section 2 in \cite{SAS} for an analogous computation.

\begin{remark}
As shown in \cite{SAS}, the  constraints  \eqref{chi=31}-\eqref{FSAS}  uniquely determine all $F_{g,n|2m}$ except $F_{g,n|2m+2}(J|0,j_0,K)$ due to existence of the extra fermionic variable $\theta^0$. However, since $F_{g,n|2m+2}(J|j_0,0,K)$ is fixed, we invole the antisymmetry of fermionic entries to fix
\begin{equation}
 F_{g,n|2m+2}(J|0,j_0,K)=-F_{g,n|2m+2}(J|j_0,0,K).
\end{equation}
This additional treatment is an analogous role to the second term in \eqref{FR}.
\end{remark}

\subsubsection{Part 2}
Our next task is to find the same set of equations \eqref{chi=31}-\eqref{FSAS} from the abstract super loop equations that respect the polarization. Note that by definition, we can always expand $\omega_{g,n|2m}$ for $g,n,m\in\mathbb{Z}_{\geq0}$ with $2g+n+2m\geq3$ in the form
\begin{equation}
\omega_{g,n|2m}(J|K)=\sum_{\substack{i_1,...,i_n\geq1\\j_1,...,j_{2m}\geq0}}\hat{F}_{g,n|2m}(J|K)\bigotimes_{k=1}^nd\xi_{-i_k}(z_k)\bigotimes_{l=1}^{2m}\eta_{-j_l}(u_l,\theta_l),\label{F hat}
\end{equation}
with some coefficients $\hat{F}_{g,n|2m}(J|K)$ which are (anti)symmetric under permutations of indices in $J$ $(K)$ respectively, but no symmetry is assumed otherwise. Therefore, we will rewrite the abstract super loop equations with respect to these coefficients $\hat{F}_{g,n|2m}(J|K)$, and show that such constraints agree with the equations obtained in \textit{Step 1}. As a consequence, uniqueness and existence in Theorem~\ref{thm:SAS1} imply that $\hat{F}_{g,n|2m}(J|K)=F_{g,n|2m}(J|K)$ which completes the proof of Theorem~\ref{thm:main} as well as Corollary~\ref{coro:main}.

Notice that the linear bosonic loop equations are equivalent to the following equations:
\begin{equation}
\forall i\in\mathbb{Z}_{\geq1},\;\;\;\;0=\underset{z=0}{\text{Res}}z^{i-1}\mathcal{L}^B_{g,n+1|2m}(z,J|K).
\end{equation}
If we substitute the expansion \eqref{F hat} in $\mathcal{L}^B_{g,n+1|2m}(z,J|K)$, it gives
\begin{equation}
\forall i\in\mathbb{Z}_{\geq1},\;\;\;\;0=\sum_{\substack{i_1,...,i_n\geq1\\j_1,...,j_{2m}\geq0}}2\hat{F}_{g,n|2m}(2i,J|K)\bigotimes_{k=1}^nd\xi_{-i_k}(z_k)\bigotimes_{l=1}^{2m}\eta_{-j_l}(u_l,\theta_l),
\end{equation}
which implies
\begin{equation}
\forall i\in\mathbb{Z}_{\geq1},\;\;\;\;\hat{F}_{g,n|2m}(2i,J|K)=0.\label{LBF}
\end{equation}
Similarly, the linear fermionic loop equations are equivalent to
\begin{equation}
\forall i\in\mathbb{Z}_{\geq0},\;\;\;\;0=\underset{z=0}{\text{Res}}\,\eta_i(z)\mathcal{L}^F_{g,n|2m}(J|z,K),
\end{equation}
which implies
\begin{equation}
\forall\,i\in\mathbb{Z}_{\geq1},\;\;\;\;\hat{F}_{g,n|2m}(J|2i-1,K)=0.\label{LFF}
\end{equation}
Therefore, \eqref{LBF} and \eqref{LFF} agree with \eqref{linear Z}.

We repeat similar procedures for the quadratic fermionic and bosonic loop equations, but computations become tedious. The quadratic fermionic loop equations are equivalent to
\begin{equation}
\forall\,i\in\mathbb{Z}_{\geq1},\;\;\;\;0=-\frac12\underset{z=0}{\text{Res}}\frac{\eta_i(z)}{z^2dz}\mathcal{Q}^{FB}_{g,n|2m}(J|z,K),\label{QFB2}
\end{equation}
where the overall $-\frac12$ factor is inserted for convention. Let us consider terms involving $\omega_{0,1|0}$ whose residues can be easily obtained as
\begin{equation}
-\frac12\sum_{p\geq2,q\geq0}(-1)^p\left(1+(-1)^i\right)\tau_p\hat{F}_{g,n|2m}(J|i+p-3,K),
\end{equation}
where we have omitted the $\bigotimes d\xi_J\bigotimes\eta_K$ factor. Notice that these terms trivially vanish if $i$ is odd. When $i$ is even, we redefine $i\rightarrow 2i$, and they become
\begin{equation}
\forall\,i\in\mathbb{Z}_{\geq1},\;\;\;\;\sum_{p\geq2}(-1)^{p-1}\tau_p\hat{F}_{g,n+1|2p}(J|2i+p-3,K)
\end{equation}
This agrees with the first term in \eqref{chi=33} and \eqref{FSAS} with the replacement of $F_{g,n|2m}$ with $\hat{F}_{g,n|2m}$.

For $2g+n+2m=3$, the rest of terms in \eqref{QFB2} becomes
\begin{align}
&-\frac12\underset{z=0}{\text{Res}}\frac{\eta_i(z)}{z^2dz}(\omega_{0,2|0}(z,z_1|)\omega_{0,0|2}(|-z,z_2)+\omega_{0,2|0}(-z,z_1|)\omega_{0,0|2}(|z,z_2))\nonumber\\
&=\sum_{j\geq1,k\geq0}d\xi_{-j}(z_1)\eta_{-k}(z_2)\frac{j}{1+\delta_{k,0}}\hat{C}_i^{-j|-k},
\end{align}
where
\begin{equation}
\hat{C}_i^{-j|-k}:=-\frac12\underset{z=0}{\text{Res}}\frac{\eta_{2i}(z)}{z^2dz}\Bigl(d\xi_{j}(z)\eta_{k}(-z)+d\xi_{j}(-z)\eta_{k}(z)\Bigr)=\delta_{i,1}\delta_{j,1}\delta_{k,0}.
\end{equation}
This is in agreement with \eqref{chi=35}, hence we have recovered \eqref{chi=33}.

For $2g+n+2m>3$, one can show after some manipulation that  \eqref{QFB2} can be written in the same form as \eqref{FSAS} with the replacement of $(F_{g,n|2m},C_i^{j|k})$ with $(\hat{F}_{g,n|2m},\hat{C}_i^{j|k})$ where
\begin{equation}
\forall\,j,k\in\mathbb{Z},\;\;\;\;\hat{C}_i^{j|k}:=-\frac12\underset{z=0}{\text{Res}}\frac{\eta_{2i}(z)}{z^2dz}\Bigl(d\xi_{-j}(z)\eta_{-k}(-z)+d\xi_{-j}(-z)\eta_{-k}(z)\Bigr).\label{C hat}
\end{equation}
Note that 
we have defined $d\xi_0=0$ for convention. One can explicitly compute the residues \eqref{C hat} and check that the result is in complete agreement with \eqref{C3}. That is,
\begin{equation}
\forall i\in\mathbb{Z}_{\geq1},\;\;\;\;\forall j,k\in\mathbb{Z},\;\;\;\;\hat{C}_i^{j|k}=C_i^{j|k}.
\end{equation} 
This implies that $\hat{F}_{g,n|2m}(J|K)$ satisfies the same equation as \eqref{FSAS}.

Similarly, the quadratic bosonic loop equations are equivalent to
\begin{equation}
\forall i \in\mathbb{Z}_{\geq1},\;\;\;\;0=-\frac12\underset{z=0}{\text{Res}}\frac{d\xi_i(z)}{z^2dzdz}\left(\mathcal{Q}^{BB}_{g,n+1|2m}(z,J|K)+\mathcal{Q}^{FF}_{g,n+1|2m}(z,J|K)\right).\label{QBB2}
\end{equation}
Almost all computations are similar to those for the quadratic fermionic loop equations, and one can manipulate the quadratic bosonic loop equations into the same form as \eqref{FSAS} with the replacement of $(F_{g,n|2m},C_i^{jk|},C_i^{|jk})$ with $(\hat{F}_{g,n|2m},\hat{C}_i^{j|k},\hat{C}_i^{|j,k})$ where 
\begin{align}
\hat{C}_i^{j,k|}&=-\frac12\underset{z=0}{\text{Res}}\frac{d\xi_{2i-1}(z)}{zdzdz}\Bigl(d\xi_{-j}(z)d\xi_{-k}(-z)+d\xi_{-j}(-z)d\xi_{-k}(z)\Bigr),\\
\hat{C}_i^{|j,k}&=-\frac14\underset{z=0}{\text{Res}}\frac{d\xi_{2i-1}(z)}{zdzdz}\Bigl(\mathcal{D}_z\cdot\eta_{-j}(z)\eta_{-k}(-z)+\mathcal{D}_z\cdot\eta_{-j}(-z)\eta_{-k}(z)\Bigr)-(j\leftrightarrow k).
\end{align}
Note that the antisymmetrization of $\hat{C}_i^{|j,k}$ is a consequence of the sign factor $(-1)^{\rho}$ in \eqref{QFF}. By explicit computations of the residues, one confirms that
\begin{equation}
\forall i\in\mathbb{Z}_{\geq1},\;\;\;\;,\forall j,k\in\mathbb{Z},\;\;\;\;\hat{C}_i^{j,k|}=C_i^{j,k|},\;\;\;\;\hat{C}_i^{|j,k}=C_i^{|j,k}
\end{equation}

The only factor that does not have any analogue in the fermionic loop equations is the one that corresponds to \eqref{D1}. This appears in the quadratic bosonic loop equation for $g=1,n=m=0$ in the form:
\begin{equation}
\hat{D}_i=-\frac12\underset{z=0}{\text{Res}}\frac{d\xi_{2i-1}(z)}{zdzdz}\left(\omega_{0,2|0}(z,-z|)-\frac12\Bigl(\mathcal{D}_z\cdot\omega_{0,0|2}(|z,u)+\mathcal{D}_u\cdot\omega_{0,0|2}(|u,z)\Bigr)\Bigr|_{u=\sigma(z)}\right).
\end{equation}
Again, one can show that $\hat{D}_i=D_i$.

In summary, $\hat{F}_{g,n|2m}$ satisfy precisely the same equations as those that $F_{g,n|2m}$ do. Thus, uniqueness of solution implies $\hat{F}_{g,n|2m}=F_{g,n|2m}$. This proves Theorem~\ref{thm:main} and Corollary~\ref{coro:main}.

\begin{remark}
For irregular cases $(\epsilon=1)$, all we have to modify from the above analysis is to shift the indices $i\rightarrow i-1$ in $H_i^2,F_i^2$, or equivalently, in terms of abstract super loop equations, we shift $i\rightarrow i-2$ in \eqref{QFB2} and \eqref{QBB2}. All other computations are completely parallel. 
\end{remark}


\subsection{Proof of Proposition~\ref{prop:reduction}}\label{sec:proof of reduction}

The case with $\tau_l=\delta_{l,3}$ is proven in \cite{Beckers,McArthur} but with a twist. For consistency, however, first we directly prove Proposition~\ref{prop:reduction} except the truncation property, and then consider truncation by consulting the arguments in \cite{Beckers,McArthur}. 

The idea of the proof is as follows. In terms of the super topological recursion, Proposition~\ref{prop:STR} implies that $\omega_{g,n|0}$ and $\omega_{g,n|2}$ are determined by themselves without the knowledge of $\omega_{g,n|2m}$ for $m\geq2$. This is because \eqref{FR} involves only $\omega_{g,n|0}$ and $\omega_{g,n|2}$. Even though $\omega_{g,n|4}$ appear if we use \eqref{BR} to compute $\omega_{g,n|2}$, the results should match with those by \eqref{FR} since existence of solution is guaranteed. An equivalent statement in terms of super Airy structures is that the leading order of $H_i^2Z=F_i^2Z=0$ with respect to Grassmann variables gives a set of constraints that uniquely determines $F_{g,n|0}$ and $F_{g,n|2}$ without the knowledge of $F_{g,n|2m\geq4}$. Higher order constraints involve $F_{g,n|2m\geq4}$. Therefore, it is sufficient for our purpose to only focus on the leading order of the differential constraints $H_i^2Z=F_i^2Z=0$.

Now we consider the super Airy structure $\mathcal{S_A}^{\tau_3}=\{ H^1_i, F_i^1, H^{2,\tau_3}_i,F^{2,\tau_3}_i\}$ with  $\tau_l=\delta_{l,3}$ $(\epsilon=3)$ and its associated free energy $F(\mathcal{S_A}^{\tau_3})$. First, we can set $x^{2i}=\theta^{2i+1}=0$  for all $i\in\mathbb{Z}_{>0}$ without loss of generality (see Section~\ref{sec:going back}). Next, let us define $\hat{F}^s,\hat{F}^s_g$ by
\begin{equation}
\hat F^s=\sum_{g\geq0}\hbar^{g-1}\hat F^s_g=\sum_{g\geq0}\hbar^{g-1}2^g\left(F_g(\mathcal{A}^{\tau_3})-\frac12\sum_{i,j\geq0}\theta^{2i}\theta^{2j}\frac{\partial^2 F_g(\mathcal{A}^{\tau_3})}{\partial x^{2i+1}\partial x^{2j-1}}\right)+\mathcal{O}(\theta)^4,\label{ansatz1}
\end{equation}
where $\mathcal{A}^{\tau_3}=\{ H^1_i,\check H^{2,\tau_3}_i\}$ is the Airy structure with $\tau_l=\delta_{l,3}, \phi_{kl}=0$ and $F(\mathcal{A}^{\tau_3})$ is the associated free energy. Then, since $\hat{F}^s$ satisfies property (1) and (2) of Theorem \ref{thm:SAS1}, it suffices to claim that $F(\mathcal{S_A}^{\tau_3})=\hat F^s+\mathcal{O}(\theta)^4$, if we show: 
\begin{equation}
e^{-\hat F^s} H_i^{2,\tau_3}e^{\hat F^s}=\mathcal{O}(\theta)^2,\;\;\;\;e^{-\hat F^s} F_i^{2,\tau_3}e^{\hat F^s}=\mathcal{O}(\theta)^3.\label{reduction2}
\end{equation}

This can be checked by directly substituting \eqref{ansatz1} into \eqref{reduction2}. First, one finds for all $i\in\mathbb{Z}_{>0}$, 
\begin{align}
e^{-\hat{F}^s}H_i^{2,\tau_3}e^{\hat{F}^s}=&\sum_{g\geq0}(2\hbar)^g\Biggl(\frac{\partial F_g(\mathcal{A}^{\tau_3})}{\partial x^{2i-1}}+\frac{x^1x^1}{2}\delta_{i,1}\delta_{g,0}+\sum_{j\geq1}(2j-1)x^{2j-1}\frac{\partial F_g(\mathcal{A}^{\tau_3})}{\partial x^{2i+2j-5}}\nonumber\\
&+\frac12\sum_{j+k=i-1}\left(\sum_{g_1+g_2=g}\frac{\partial F_{g_1}(\mathcal{A}^{\tau_3})}{\partial x^{2j-1}}\frac{\partial F_{g_2}(\mathcal{A}^{\tau_3})}{\partial x^{2k-1}}+\frac{\partial^2F_{g-1}(\mathcal{A}^{\tau_3})}{\partial x^{2j-1}\partial x^{2k-1}}\right)\nonumber\\
&+\frac{\hbar}{8}\delta_{i,2}\delta_{g,1}\Biggr)+\mathcal{O}(\theta)^2,
\end{align}
The leading order in $\theta$ vanishes thanks to the assumption that $ F(\mathcal{A}^{\tau_3})$ is the free energy of the Airy structure $\mathcal{A}^{\tau_3}$. Next, one can show that
\begin{equation}
e^{-\hat{F}^s}F^{2,\tau_3}_ie^{\hat{F}^s}=\sum_{j\geq0}\theta^{2j}\frac{\partial}{\partial x^{2j+1}}\left(\frac12e^{-\hat{F}^s}H_{i}^{2,\tau_3}e^{\hat{F}^s}\right)-\sum_{j\geq0}\theta^{2j}\frac{\partial}{\partial x^{2j-1}}\left(\frac12e^{-\hat{F}^s}H_{i+1}^{2,\tau_3}e^{\hat{F}^s}\right)+\mathcal{O}(\zeta)^3.
\end{equation}
Thus, the leading order in $\theta$ again vanishes thanks to the assumption. We can apply the same technique to the super Airy structure $\mathcal{S_A}^{\tau_1}$ with $\tau_l=\delta_{l,1}$ $(\epsilon=1)$. 
This shows that \eqref{reduction} holds, at least for $\mathcal{S_A}^{\tau_1},\mathcal{S_A}^{\tau_3}$. 

We will now generalize the above results for an Airy structure $\mathcal{A}^{\tau}$ with arbitrary dilaton shift $(\tau_l)_{l>0}$. It is easy to show by induction in $n\in\mathbb{Z}_{\geq0}$ that every nonzero coefficient $F_{0,n+3}(\mathcal{A}^{\tau_3})(i_1,i_2,i_3,I)$ is always in the form:
 \begin{equation}
F_{0,n+3}(\mathcal{A}^{\tau_3})(1,1,1,I)\label{reduction3}
\end{equation}
where  $I=\{i_4,...,i_{n+3}\}$. That is, at least three of the entries must be 1. Recall that the free energy associated with any Airy structure is unique up to addition of terms in $\hbar^{-1}\mathbb{C}[[\hbar]]$. Importantly, \eqref{Phi on modes} shows that taking conjugate with respect to arbitrary $\tau_l$ merely shifts each variable $x^l\mapsto x^l+\tau_l$. Thus, if we define $\bar F$ by conjugation
\begin{equation}
e^{\bar F}=\exp\left(\sum_{l\geq2}\frac{\tau_l}{l}J_l\right)e^{F(\mathcal{A}^{\tau_3})}\exp\left(-\sum_{l\geq2}\frac{\tau_l}{l}J_l\right),\label{reduction4}
\end{equation}
\eqref{reduction3} ensures that no unwanted terms such as $\bar F_{0,1}$ and $\bar F_{0,2}$ appear in $\bar F$. Then by construction, all differential operators in $\mathcal{A}^{\tau}$ annihilate \eqref{reduction4}, hence it follows that
\begin{equation}
\bar F-F(\mathcal{A}^{\tau})\in\hbar^{-1}\mathbb{C}[[\hbar]]\label{reduction5}
\end{equation}

With this in mind, let us define $\bar F^s$ as
\begin{equation}
e^{\bar F^s}=\exp\left(\sum_{l\geq2}\frac{\tau_l}{l}J_l\right)e^{F(\mathcal{S_A}^{\tau_3})}\exp\left(-\sum_{l\geq2}\frac{\tau_l}{l}J_l\right).
\end{equation}
Then as a consequence of \eqref{reduction5}, order by order in $\hbar$, we have
\begin{equation}
\bar F_g^s= 2^g\left(c_g+F_g(\mathcal{A}^{\tau})-\frac12\sum_{i,j\geq0}\theta^i\theta^j\frac{\partial^2F_g(\mathcal{A}^{\tau})}{\partial x^i\partial x^{j-1}}\right)+\mathcal{O}(\theta^4),
\end{equation}
for some $c_g\in\mathbb{C}$. Importantly, $\bar F_0^s$ does not have the unwanted terms, hence $\bar F^s$ satisfies  property (1) and (2) of Theorem~\ref{thm:SAS1}.
 
Finally, let  $\mathcal{S_A}^{\tau}=\{H^1_i,F^1_i,H^{2,\tau}_i,F^{2,\tau}_i\}$ be a set of differential operators of arbitrary dilaton shift with $\epsilon=3$. Explicitly,
\begin{align}
H_i^{2,\tau}&=\exp\left(\sum_{l\geq2}\frac{\tau_l}{l}J_l\right)H_i^{2,\tau_3}\exp\left(-\sum_{l\geq2}\frac{\tau_l}{l}J_l\right),\\
F_i^{2,\tau}&=\exp\left(\sum_{l\geq2}\frac{\tau_l}{l}J_l\right)F_i^{2,\tau_3}\exp\left(-\sum_{l\geq2}\frac{\tau_l}{l}J_l\right).
\end{align}
Thus, by construction (recall \eqref{reduction2}), we find that $\bar F^s$ satisfies 
\begin{equation}
e^{-\bar F^s}H_i^{2,\tau}e^{\bar F^s}=\mathcal{O}(\theta)^2,\;\;\;\;e^{-\bar F^s}F_i^{2,\tau}e^{\bar F^s}=\mathcal{O}(\theta)^3.
\end{equation}
Then, it follows that
\begin{equation}
F(\mathcal{S_A}^{\tau})=\bar F^s +\mathcal{O}(\theta^4).
\end{equation}
Again, one can repeat the same trick for every set $\mathcal{S_A}^{\tau}$ of differential operators of arbitrary dilaton shift $\epsilon=1$. This proves that \eqref{reduction} holds for any $\mathcal{S_A}^{\tau}$.

\subsubsection{Justification of Proposition~\ref{prop:SKdV}}\label{sec:just:SKdV}
We finish the proof of Proposition~\ref{prop:reduction} with comments about truncation phenomena which simultaneously justifies Proposition~\ref{prop:SKdV}. It is shown in \cite{Beckers} that the partition function $e^{\mathcal{F}_S}$ of the continuum limit of supereigenvalue models is annihilated by  $(G_{n+1/2})_{n\geq-1}$ defined in Eq. (29) in \cite{Beckers} which are represented by a set of variables $(t_i,\tau_i)_{i\geq0}$. Note that $(\tau_i)_{i\geq0}$ denote their Grassmann variables only here and in \eqref{dictionary} below, but they denote dilaton shift parameters anywhere else. Furthermore, they conjectured that $\mathcal{F}_S$ is given by the partition function of the continuum limit of Hermitian matrix models which also relates to $F_g(\mathcal{A}^{\tau_3})$ \cite{Dijkgraaf,K,W}. This conjecture, in particular, its truncation property is proven in \cite{McArthur}. Explicitly, their notation is translated into ours as follows:
\begin{equation}
\forall i\geq0,\;\;\;\;t_i=x^{2i+1},\;\;\;\;\tau_i=\frac{\theta^{2i}}{\sqrt{2}},\;\;\;\;x^{2i}=\theta^{2i+1}=0,\;\;\;\;\kappa^2=2\hbar,\label{dictionary}
\end{equation}
\begin{align}
G_{i-3/2} &= \frac{1}{\sqrt{2}\hbar}\exp\left(-\frac13J_3\right) F_i^2\exp\left(\frac13J_3\right) ,\\
e^{\mathcal{F}_S}&=\exp\left(-\frac13J_3\right)  e^{F(\mathcal{S_A}^{\tau_3})} \exp\left(\frac13J_3\right),\label{dictionary2}
\end{align}
where the left hand sides of these equalities are notation used in \cite{Beckers} and the right hand sides ours. Note that  Eq. (31) in \cite{Beckers} is obtained by \eqref{reduction} in terms of the $\kappa^2$-expansion instead of $\hbar$. Since the degree of Grassmann variable dependence never increases under taking conjugate with respect to $(\tau_l)_{l>0}$, we conclude from \eqref{dictionary2} and from the results in \cite{Beckers,McArthur} that $F(\mathcal{S_A}^{\tau})$  with $\epsilon=3$ truncates at quadratic order in Grassmann variables $\theta^i$. We cannot apply this argument to cases with $\epsilon=1$ because the results in \cite{Beckers,McArthur} are valid only for $\tau_l=\delta_{l,3}$. This completes the proof of Proposition~\ref{prop:reduction}.

With \eqref{dictionary2} being shown, one can easily follow \cite{SL1,SL3} to compute  correlation functions of $(2,4\ell)$-minimal superconformal models coupled to Liouville supergravity. Note that $\mathcal{F}_S$ has nonzero $F_{0,1|0}$, $F_{0,2|0}$, and $F_{0,0|2}$ unlike $F(\mathcal{S_A}^{\tau_3})$ due to the (inverse) dilaton shift \eqref{dictionary2}.

\subsection{Proof of Proposition~\ref{prop:NS}}\label{sec:NS}

We apply the $\mathcal{N}=1$ super topological recursion to prove  Proposition~\ref{prop:NS} instead of super Airy structures. Concretely, we first derive the super loop equations of supereigenvalue models in the Neveu-Schwarz sector, and then next, we determine an appropriate local super spectral curve with the help of the results shown in \cite{BO}. Finally, we evaluate the super loop equations on the local super spectral curve and show that the super loop equations of supereigenvalue models fit into the framework of abstract super loop equations. \cite{BO} shows a recursive formula for correlation functions without half-order differentials thanks to a great simplification due to \cite{Beckers,McArthur}. Here instead, we will take a different definition of correlation functions in such a way that the $\mathcal{N}=1$ super topological recursion suits well. Parts of computations and analyses below are taken from \cite{BO}.

\subsubsection{Super Loop Equations}
As the first step towards the proof of Proposition~\ref{prop:NS}, we derive the super loop equations of supereigenvalue models. Let $F_{NS}=\log Z_{NS}$ be the free energy of 1-cut supereigenvalue models in the Neveu-Schwarz sector with coupling constants $\{g_k,\xi_{k+\frac12}\}_{k\in\mathbb{Z}_{\geq0}}$. The bosonic and fermionic potentials are defined with the coupling constants and a formal variable $x$ as
\begin{equation}
V(x)=\sum_{k\geq0}g_kx^k,\;\;\;\;\Psi_{NS}(x)=\sum_{k\geq0}\xi_{k+\frac12}x^k.
\end{equation}
The partition function is annihilated by super Virasoro operators $\{L_n,G_{n+\frac12}\}_{n\in\mathbb{Z}_{\geq-1}}$ in the Neveu-Schwarz sector,
\begin{equation}
\forall\,n\in\mathbb{Z}_{\geq-1},\;\;\;\;L_nZ_{NS}=G_{n+\frac12}Z=0,
\end{equation}
where the representation of these operators can be found in \cite{BO}. Note that $F_{NS}$ is \emph{not} the free energy of the associated super Airy structure. Accordingly, these super Virasoro operators do \emph{not} form a nontrivial super Airy structure\footnote{After an appropriate conjugation, they can form a super Airy structure whose associated free energy is zero. See \cite{SAS}.}.

Let $2N$ be the number of bosonic (equiv. fermionic) eigenvalues in supereigenvalue models. It can be shown that the free energy enjoy the $1/N$-expansion, that is,
\begin{equation}
F_{NS}=\sum_{g\geq0}\left(\frac{1}{N}\right)^{2-2g}F_{NS,g}.\label{1/N}
\end{equation}
We introduce the bosonic and fermionic loop insertion operators as
\begin{equation}
\frac{\partial}{\partial V(x)}:=-\sum_{k\geq0}\frac{1}{x^{k+1}}\frac{\partial}{\partial g_k},\;\;\;\;\frac{\partial}{\partial \Psi_{NS}(x)}:=-\sum_{k\geq0}\frac{1}{x^{k+1}}\frac{\partial}{\partial \xi_{k+\frac12}}.
\end{equation}
Then, correlation functions $W_{g,n|2m}$ are defined by
\begin{equation}
W_{g,n|m}(J|K)=\left(\frac{1}{N}\right)^{2g-2+n+m}\prod_{i=1}^n\frac{\partial}{\partial V(x_i)}\prod_{i=0}^{m-1}\frac{\partial}{\partial \Psi_{NS}(\tilde{x}_{m-i})}F_{NS,g}.
\end{equation}
Note that the power of $N$ is inserted so that $W_{g,n|m}(J|K)$ are independent of $N$. Also, the ordering of the fermionic loop insertion operators is important as sign may appear if the order is chosen differently. In particular,  our definition is different from that in \cite{BO,O} to match with the abstract super loop equations.

Their super loop equations are derived from the following series
\begin{equation}
\sum_{n\geq-1}\frac{1}{x^{n+2}}\frac{1}{Z_{NS}}L_nZ_{NS}=0,\;\;\;\;\sum_{n\geq-1}\frac{1}{x^{n+2}}\frac{1}{Z_{NS}}G_{n+\frac12}Z_{NS}=0.
\end{equation}
As shown in \cite[Section 3.4]{BO}, our first step is to manipulate the above series into the following forms:
\begin{align}
P^{NS,BB}_{g,1|0}(x|)=&-V'(x)W_{g,1|0}(x|)+\frac12\sum_{g_1+g_2=g}W_{g_1,1|0}(x|)W_{g_2,1|0}(x|)+\frac12W_{g-1,2|0}(x,x|)\nonumber\\
&-\frac12W'_{g,0|1}(|x)\Psi_{NS}(x)-\frac12\Psi'_{NS}(x)W_{g,0|1}(|x)\nonumber\\
&+\frac12\sum_{g_1+g_2=g}W'_{g_1,0|1}(|x)W_{g_2,0|1}(|x)-\frac12W'_{g_2,0|1}(|x,\tilde{x})\Bigr|_{\tilde{x}=x},\label{NSBB1}\\
P^{NS,FB}_{g,0|1}(|x)=&-V'(x)W_{g,0|1}(|x)-\Psi_{NS}(x)W_{g,1|0}(x|)\nonumber\\
&+\sum_{g_1+g_2=g}W_{g_1,1|0}(x|)W_{g_2,0|1}(|x))+W_{g-1,1|1}(x|x),\label{NSFB1}
\end{align}
where the prime denotes the derivative with respect to $x$ and
\begin{align}
P^{NS,BB}_{g,1|0}(x|)&=\sum_{n,k\geq0}x^n\left((n+k+2)g_{n+k+2}\frac{\partial}{\partial g_k}+\frac12(n+2k+3)\xi_{n+k+\frac52}\frac{\partial}{\partial \xi_{k+\frac12}}\right)F_{NS,g},\label{P1}\\
P^{NS,FB}_{g,0|1}(|x)&=\sum_{n,k\geq0}x^n\left((n+k+2)g_{n+k+2}\frac{\partial}{\partial \xi_{k+\frac12}}+\xi_{n+k+\frac32}\frac{\partial}{\partial g_k}\right)F_{NS,g}.\label{P2}
\end{align}

If we act an arbitrary number of times with the loop insertion operators on $P^{NS,BB}_{g,1|0}(x|)$ and $P^{NS,FB}_{g,0|1}(|x)$, we get from \eqref{P1}, \eqref{P2} that
\begin{align}
Q^{NS,BB}_{g,n+1,m}(x,J|K):=&\left(\frac{1}{N}\right)^{n+m}\prod_{i=1}^n\frac{\partial}{\partial V(x_i)}\prod_{i=0}^{m-1}\frac{\partial}{\partial \Psi_{NS}(\tilde{x}_{m-i})}P^{NS,BB}_{g,1|0}(x|)\nonumber\\
=&P^{NS,BB}_{g,n+1|m}(x,J|K)-\sum_{i=1}^n\frac{d}{dx_i}\frac{W_{g,n|m}(J|K)}{x-x_i}\nonumber\\
&+\sum_{j=1}^m(-1)^m\left(\frac12\frac{d}{dx}\frac{W_{g,n|m}(J|K)}{x-\tilde{x}_j}-\frac{d}{d\tilde x_j}\frac{W_{g,n|m}(J|K)}{x-\tilde{x}_j}\right),\label{NS2}\\
Q^{NS,FB}_{g,n|m+1}(J|x,K):=&\left(\frac{1}{N}\right)^{n+m}\prod_{i=1}^n\frac{\partial}{\partial V(x_i)}\prod_{i=0}^{m-1}\frac{\partial}{\partial \Psi_{NS}(\tilde{x}_{m-i})}P^{NS,BF}_{g,1|0}(x|)\nonumber\\
=&P^{NS,FB}_{g,n|m+1}(J|x,K)-\sum_{i=1}^n\frac{d}{dx_i}\frac{W_{g,n-1|m+1}(J\backslash x_i|x_i,K)}{x-x_i}\nonumber\\
&-\sum_{j=1}^m(-1)^m\frac12\frac{W_{g,n|m}(\tilde{x}_j,J|K\backslash \tilde{x}_j)}{x-\tilde{x}_j}\label{NS4}
\end{align}
where
\begin{align}
P^{NS,BB}_{g,n+1|m}(x,J|K)=&\left(\frac{1}{N}\right)^{n+m}\sum_{n,k\geq0}x^n\left((n+k+2)g_{n+k+2}\frac{\partial}{\partial g_k}+\frac12(n+2k+3)\xi_{n+k+\frac52}\frac{\partial}{\partial \xi_{k+\frac12}}\right)\nonumber\\
&\cdot\prod_{i=1}^n\frac{\partial}{\partial V(x_i)}\prod_{i=0}^{m-1}\frac{\partial}{\partial \Psi_{NS}(\tilde{x}_{m-i})}F_{NS,g},\label{NS3}
\end{align}
\begin{align}
P^{NS,FB}_{g,n|m+1}(x,J|K)=&\left(\frac{1}{N}\right)^{n+m}\sum_{n,k\geq0}x^n\left((n+k+2)g_{n+k+2}\frac{\partial}{\partial \xi_{k+\frac12}}+\xi_{n+k+\frac32}\frac{\partial}{\partial g_k}\right)\nonumber\\
&\cdot\prod_{i=1}^n\frac{\partial}{\partial V(x_i)}\prod_{i=0}^{m-1}\frac{\partial}{\partial \Psi_{NS}(\tilde{x}_{m-i})}F_{NS,g}.\label{NS5}
\end{align}
At the same time, \eqref{NSBB1} and \eqref{NSFB1} imply that
\begin{align}
Q^{NS,BB}_{g,n+1,m}(x,J|K)=&-V'(x)W_{g,n+1|m}(x,J|K)+\sum_{i=1}^n\frac{1}{(x-x_i)^2}W_{g,n,m}(x,J\backslash x_i|K)\nonumber\\
&+\frac12W_{g-1,n+2|m}(x,x,J|K)-\frac12W'_{g-1,n|m+2}(J|x,\tilde{x},K)\Bigr|_{\tilde{x}=x}\nonumber\\
&+\frac12\sum_{g_1+g_2=g}\sum_{\substack{J_1\cup J_2=J \\ K_1\cup K_2=K}}(-1)^{\rho}W_{g_1,n_1+1|2m_1}(x,J_1|K_1)W_{g_2,n_2+1|m_2}(x,J_2|K_2)\nonumber\\
&-\frac12W'_{g,n|m+1}(J|x,K)\Psi_{NS}(x)+\frac12\sum_{j=1}^m(-1)^{j-1}\frac{1}{x-\tilde{x}_j}W'_{g,n|m}(J|x,K\backslash \tilde{x}_j)\nonumber\\
&-\frac12\Psi'_{NS}(x)W_{g,n|m+1}(J|x,K)-\frac12\sum_{j=1}^m(-1)^{j-1}\frac{d}{dx}\frac{1}{x-\tilde{x}_j}W_{g,n|m}(J|x,K\backslash \tilde{x}_j)\nonumber\\\nonumber\\
&+\frac12\sum_{g_1+g_2=g}\sum_{\substack{J_1\cup J_2=J \\ K_1\cup K_2=K}}(-1)^{\rho}W'_{g_1,n_1|m_1+1}(J_1|x,K_1)W_{g_2,0|1}(J_2|x,K_2),\label{NSBB}
\end{align}
and
\begin{align}
Q^{NS,FB}_{g,n|m+1}(J|x,K)=&-V'(x)W_{g,n|m+1}(J|x,K)+\sum_{i=1}^n\frac{1}{(x-x_i)^2}W_{g,n-1,m+1}(J\backslash x_i|x,K)\nonumber\\
&-\Psi_{NS}(x)W_{g,n+1|,}(x,J|K)-\sum_{j=1}^m(-1)^{j-1}\frac{1}{x-\tilde{x}_j}W_{g,n+1|m-1}(x,J|K\backslash \tilde{x}_j)\nonumber\\
&+W_{g-1,n+1|m+1}(x,J|x,K)\nonumber\\
&+\sum_{g_1+g_2=g}\sum_{\substack{J_1\cup J_2=J \\ K_1\cup K_2=K}}(-1)^{\rho}W_{g_1,n_1+1|2m_1}(x,J_1|K_1)W_{g_2,n_2|m_2+1}(J_2|x,K_2).\label{NSFB}
\end{align}
In the computations, we used
\begin{equation}
\frac{\partial}{\partial V(x_1)}V(x)=-\frac{1}{(x-x_1)^2},\;\;\;\;\frac{\partial}{\partial \Psi_{NS}(\tilde{x}_1)}\Psi_{NS}(x)=\frac{1}{x-\tilde{x}_1}.\label{NS1}
\end{equation}
We note that these super loop equations are derived without referring to the reduction \cite{Beckers,McArthur}. Although \eqref{NSBB} and \eqref{NSFB} are called the super loop equations of supereigenvalue models in the Neveu-Schwarz sector, we still have to show that they are examples of \emph{abstract} super loop equations, when evaluated on an appropriate local super spectral curve.

\subsubsection{Local Super Spectral Curve}

The second step is to find an appropriate local super spectral curve. To do so, however, let us start with a global picture. We set the potentials $V(x),\Psi(x)$ to be polynomials. Then, \cite{BO} showed that if we consider a hyperelliptic curve
 \begin{equation}
y^2-(x-1)(x+1)(M(x))^2=0,
\end{equation}
with parametrization
\begin{equation}
x(z)=\frac12\left(u+\frac{1}{u}\right),\;\;\;\;y(z)=\frac12\left(u-\frac{1}{u}\right)M(x(u)),
\end{equation}
 we find
\begin{equation}
\omega_{0,1|0}(u|):=M(x)y(u)dx(u)=\frac12\left(W_{0,1|0}^{(0)}(u|)-V'(x(u))\right)dx(u),\label{NSw010}
\end{equation}
\begin{equation}
\omega_{0,2|0}(u_1,u_2|):=\frac12W^{(0)}_{0,2|0}(u_1,u_2|)dx_1dx_2+\frac{dx_1dx_2}{(x_1-x_2)^2}=\frac{du_1du_2}{(u_1-u_2)^2},\label{NSw0201}
\end{equation}
\begin{equation}
W_{0,0|2}^{(0)}(|u_1,u_2)=-(x_1-x_2)W^{(0)}_{0,2|0}(u_1,u_2|),
\end{equation}
where the superscript $(0)$ denotes the term independent of $\xi_k$-couplings. Note that they are globally well-defined on the hyperelliptic curve.

Next, we will move to a local picture and define $\omega_{0,0|2}$. There are two ramification points, $u=\pm1$, and we focus on one of them, $u=1$ -- the case with $u=-1$ goes parallel. Moving from a global picture to a local picture is done by considering a local coordinate $z$ satisfying
\begin{equation}
x=\frac12\left(u+\frac{1}{u}\right)=1+\frac{z^2}{2},\label{local coordinate}
\end{equation}
and by rewriting everything as formal expansion in $z$. Note that \eqref{local coordinate} is a valid equation only in the neighbourhood of the ramification point $u=1$. Furthermore, we extend this local patch $\mathbb{C}$ to $\mathbb{C}^{1|1}$ together with a Grassmann variable $\theta$. With this setting, we have from \eqref{NSw010} and \eqref{NSw0201} that
\begin{align}
\omega_{0,1|0}(z|)&=\frac12M(x)\left(u(z)-\frac{1}{u(z)}\right)z\,dz,\\
\omega_{0,2|0}(z_1,z_2|)&=\frac{1}{(u(z_1)-u(z_2))^2}\frac{du(z_1)}{dz_1}\frac{du(z_2)}{dz_2}dz_1 dz_2,
\end{align}
where they should be understood as formal expansion in $z,z_1,z_2$. Also, we define
\begin{align}
\omega_{0,0|2}(|z_1,z_2):&=\left(\frac12W^{(0)}_{0,2|0}(|z_1,z_2)-\frac{1}{(x_1-x_2)}\right)\Theta_1\Theta_2\nonumber\\
&=-\frac{x(z_1)-x(z_2)}{dx(z_1)dx(z_2)}\omega_{0,2|0}(z_1,z_2|)\Theta_1\Theta_2
\end{align}
Again this should be understood as formal expansion in $z_1,z_2$.  Notice that $(\omega_{0,1|0}, \omega_{0,2|0}, \omega_{0,0|2})$ in formal expansion in $z$ provide the defining data of one of the two components of a local super spectral curve (Definition~\ref{def:SCmulti}). Therefore, together with a similar analysis for the case with $u=-1$ and mixed cases for $ \omega_{0,2|0}, \omega_{0,0|2}$, we have found an appropriate local super spectral curve of two components given in \eqref{tauNS}, \eqref{phiNS}, and \eqref{psiNS}.

\subsubsection{Abstract Super Loop Equations}
The final task towards the proof of Proposition~\ref{prop:NS} is to transform the super loop equations of supereigenvalue models evaluated on the above local super spectral curve into the abstract super loop equations on the local supoer spectral curve. Again we only discuss the case with $u=1$ and the case with $u=-1$ immediately follow. By construction, we know that $\omega_{0,1|0},\omega_{0,2|0}, \omega_{0,0|2}$ satisfy the linear abstract super loop equations
\begin{align}
\omega_{0,1|0}(z_1|)+\omega_{0,1|0}(\sigma(z_1)|)&=0,\label{NSw0101}\\
\omega_{0,2|0}(z_1,z_2|)+\omega_{0,2|0}(\sigma(z_1),z_2|)&=\frac{dx_1dx_2}{(x_1-x_2)^2},\label{NSw020}\\
\omega_{0,0|2}(|z_1,z_2)+\omega_{0,0|2}(|\sigma(z_1),z_2)&=-\frac{\Theta_1\Theta_2}{x_1-x_2}.\label{NSw002}
\end{align}
For $2g+n+2m\geq3$, we define
\begin{equation}
\omega_{g,n|2m}(J|K)=2^{g-1}W^{(0)}_{g,n|2m}(J|K)\bigotimes_{i=1}^ndx(z_i)\bigotimes_{j=1}^{2m}\Theta(\tilde{z}_j).\label{wNS}
\end{equation}
The $2^{g-1}$ is inserted for convention, but this is indeed related to \eqref{dictionary} -- one can think of $\kappa=1/N$. We would like to rewrite the $\xi_k$-coupling independent terms on the right hand sides of \eqref{NSBB} and \eqref{NSFB} with respect to $\omega_{g,n|2m}$. In order to do so, let us give a list of what we have to do:
\begin{enumerate}
\item multiply to \eqref{NSBB} 
\begin{equation}
dx(z)^{\otimes2}\bigotimes_{i=1}^ndx(z_i)\bigotimes_{j=1}^{2m}\Theta(\tilde{z}_j)
\end{equation}
\item multiply to \eqref{NSFB} 
\begin{equation}
dx(z)\Theta(z)\bigotimes_{i=1}^ndx(z_i)\bigotimes_{j=1}^{2m-1}\Theta(\tilde{z}_j)
\end{equation}
\item use \eqref{NSw010}, \eqref{NSw0101}, \eqref{NSw020}, and \eqref{NSw002} whenever possible to get rid of the following terms:
\begin{equation}
V'(x(z))dx(z),\;\;\;\;,\frac{dx(z)dx(z_i)}{(x(z)-x(z_i))^2},\;\;\;\;-\frac{\Theta(z)\Theta(\tilde x_j)}{x-\tilde x_j}.
\end{equation}
\item use the following relation whenever necessary,
\begin{equation}
\frac12dxdxW'_{0,0|2}(|z,\tilde{z})\Bigr|_{\tilde{z}=z}=\mathcal{D}_z\cdot\omega_{0,0|2}(|z,\tilde{z})\Bigr|_{\tilde{z}=\sigma(z)}=\mathcal{D}_z\cdot\omega_{0,0|2}(|\sigma(z),\tilde{z})\Bigr|_{\tilde{z}=z},
\end{equation}
\end{enumerate}

There is one more important process to arrive at the abstract super loop equations. A key observation is that after setting $V(x),\Psi(x)$ to be polynomials in $x$, both $Q^{NS,BB}_{g,n+1,m}(x,J|K)$ and $Q^{NS,FB}_{g,n|m+1}(J|x,K)$ become functions of $x$ which is regular in the neighbourhood of $z=0$. This can be explicitly seen by \eqref{NS2} and \eqref{NS4}. Thus, \emph{locally} we have
\begin{align}
Q^{NS,BB}_{g,n+1|m}(x,J|K)dxdx&\in zV^{B+}\otimes  zV^{B+},\\
Q^{NS,FB}_{g,n|m+1}(J|x,K)dx\Theta_z&\in zV_z^{B+}\otimes V_z^{F+},
\end{align}
which agree with the defining condition for the quadratic abstract super loop equations. Moreover, since it is invariant under $z\rightarrow\sigma(z)=-z$, we can show\footnote{See Section 2.3.2, \cite{BO} for an analogous analysis on a global spectral curve. Computations are parallel.} by induction in $2g+n+2m\geq3$,
\begin{align}
\omega_{g,n|2m}(z,J|K)+\omega_{g,n|2m}(\sigma(z),J|K)&=0,\label{LB1}\\
\omega_{g,n|2m}(J|z,K)+\omega_{g,n|2m}(J|\sigma(z),K)&=0.\label{LF1}
\end{align}
They agree with the linear super loop equations. With these relations in hands, we add another step to the list:
\begin{enumerate}
\item[(5)] use \eqref{LB1} and \eqref{LF1} whenever appropriate to obtain the quadratic abstract super loop equations.
\end{enumerate}
If we proceed (1) - (5) inductively,  we can show that the right hand sides of \eqref{NSBB} and \eqref{NSFB} agree with the quadratic abstract super loop equations with an overall factor of $2^{-g}$.

As the final remark, it is not \emph{a priori} guaranteed that $\omega_{g,n|2m}$ respects the polarization:
\begin{equation}
	\omega_{g,n|2m}\in \left(\bigotimes_{j=1}^nV_{z_j}^{B-}\right)\otimes \left(\bigotimes_{k=1}^{2m} V_{u_k,\theta_k}^{F\,0,-} \right).\label{wNS1}
\end{equation}
This is rather a property that we have to show by investigating their pole structures. Fortunately, this property has been shown in \cite{BO} from a global point of view, and it s straightforward to transform their results into our local description.

In summary, we have found a local super spectral curve of supereigenvalue models in the Neveu-Schwarz sector, and have shown that their correlation functions respect the polarization and that they satisfy the abstract super loop equations. Therefore, thanks to Proposition~\ref{prop:STR} and Corollary~\ref{coro:main}, the $\mathcal{N}=1$ super topological recursion uniquely constructs all correlation functions of supereigenvalue models in the Neveu-Schwarz sector. This completes the proof of Proposition~\ref{prop:NS}.

\subsection{Proof of Proposition~\ref{prop:R}}\label{sec:R}

The recursive formula for correlation functions of supereigenvalue models in the Ramond sector was recently obtained in \cite{O} where correlation functions are defined as meromorphic differentials on a certain hyperelliptic curve without a superconformal structure. Thus, similar to the NS sector, we will define correlation functions differently from how \cite{O} does so that they fit to the framework of the $\mathcal{N}=1$ super topological recursion. Since the strategy for the proof is very similar to that for the Neveu-Schwarz sector, we only point out some important differences and omit all other straightforward tediuos computations.

\subsubsection{Super Loop Equations}
We first derive the super loop equations of supereigenvalue models in the Ramond sector. Let $Z_R$ be the partition function of supereigenvalue models in the Ramond sector, then it is crucial to remark that $Z$ is annihilated by $L_{n+1},G_m$ for nonnegative integers $n,m$, and it is not annihilated by $L_0$ but rather 
\begin{equation}
L_0Z_R=\frac{1}{16}Z_R,
\end{equation}
where the explicit representation of these operators can be found in \cite{O}. In particular, $L_{-1}Z_R\neq0$. The bosonic potential $V(x)$ and the bosonic loop insertion operator are defined in the same way as in the Neveu-Schwarz sector. On the other hand, the fermionic ones are defined with half-integer powers of $x$ as 
\begin{equation}
\Psi_R(x)=\sum_{k\geq0}\xi_kx^{k-\frac12},\;\;\;\;\frac{\partial}{\partial \Psi_R(x)}=-\sum_{k\geq0}\frac{1}{1+\delta_{k,0}}\frac{1}{x^{k+\frac12}}\frac{\partial}{\partial \xi_k}.\label{FloopR}
\end{equation}
Note that the $\delta_{k,0}$ factor is due to existence of the fermionic zero mode $\psi_0$ with $\{\psi_0,\psi_0\}=1$ which is represented as
\begin{equation}
\psi_0=\xi_0+\frac12\frac{\partial}{\partial \xi_0}.
\end{equation}
Accordingly, correlation functions are defined by acing an arbitrary number of times with the loop insertion operators on the free energy $F_R=\log Z_R$ as
\begin{equation}
W_{g,n|m}(J|K)=\left(\frac{1}{N}\right)^{2g-2+n+m}\prod_{i=1}^n\frac{\partial}{\partial V(x_i)}\prod_{i=0}^{m-1}\frac{\partial}{\partial \Psi_{R}(\tilde{x}_{m-i})}F_{R,g}.
\end{equation}
In contrast to the Neveu-Schwarz sector, however, the $1/N$-expansion is an \emph{assumption} rather than a consequence due to the lack of relation to Hermitian matrix models.

It is discussed in \cite{O} that $\sqrt{x}$ is not a well-defined function on a Riemann surface. In order to avoid this issue, \cite{O} introduces a variant of the fermionic potential and fermionic loop insertion operator in order to make correlation functions well-defined on a \emph{global} spectral curve. Since our formalism in the present paper is \emph{local}, however, we can take the definition \eqref{FloopR}, which in fact seem more natural from a vertex operator algebra point of view. Effectively, one needs to divide by $\sqrt{x}$ or multiply $\sqrt{x}$ to the fermionic potential and the fermionic loop insertion operator if one wants to mathc with \cite{O}'s notation. 

The super  loop equations in the Ramond sector are derived from the following series:
\begin{equation}
\sum_{n\geq0}\frac{1}{x^{n+2}}\frac{1}{Z_R}\left(L_n-\frac{\delta_{n,0}}{16}\right)Z_R=0,\;\;\;\;\sum_{n\geq0}\frac{1}{x^{n+\frac32}}\frac{1}{Z_R}G_nZ_R=0.
\end{equation}
By acting  an arbitrary number of times with the loop insertion operators, one can manipulate and bring the two power series into similar expressions to \eqref{NSBB}, \eqref{NS2}, \eqref{NSFB}, and \eqref{NS4}. Even though computations are tedious, the results in the Ramond sector are obtained by the following replacements. See Appendix A, \cite{O} for a justification. Almost all computations are parallel:

\subsection*{NS-R dictionary}
\begin{description}
\item[NS-R 1] replace $\Psi_{NS}(x)$ with $\Psi_R(x)$
\item[NS-R 2] replace the fermionic loop insertion operator
\begin{equation}
\frac{\partial}{\partial\Psi_{NS}(x)}\mapsto\frac{\partial}{\partial\Psi_R(x)}
\end{equation}
\item[NS-R 3]  replace as follows on the right hand sides of \eqref{NSBB} and \eqref{NSFB}
\begin{equation}
\frac{1}{x-\tilde{x}_j}\mapsto\frac12\frac{x+\tilde{x}_j}{x-\tilde{x}_j}\frac{1}{\sqrt{x}\sqrt{\tilde{x}_j}},\label{R1}
\end{equation}
\item[NS-R 4] replace $(P^{NS,BB}_{g,n+1|m},P^{NS,FB}_{g,n|m+1})$ with $(P^{R,BB}_{g,n+1|m},P^{R,FB}_{g,n|m+1})$ where
\begin{align}
P^{R,BB}_{g,n+1|m}(x,J|K)=&\sum_{n\geq-1,k\geq0}x^n\left((n+k+2)g_{n+k+2}\frac{\partial}{\partial g_k}+\frac12\frac{n+2k+2}{1+\delta_{k,0}}\xi_{n+k+2}\frac{\partial}{\partial \xi_k}\right)\nonumber\\
&\times\prod_{i=1}^n\frac{\partial}{\partial V(x_i)}\prod_{i=0}^{m-1}\frac{\partial}{\partial \Psi_R(\tilde{x}_{m-i})}F_{R,g},\\
P^{R,FB}_{g,n|m+1}(J|x,K)=&\sum_{n,k\geq0}x^{n-\frac12}\left(\frac{n+k+1}{1+\delta_{k,0}}g_{n+k+1}\frac{\partial}{\partial \xi_k}+\xi_{n+k+1}\frac{\partial}{\partial g_k}\right)\nonumber\\
&\times\prod_{i=1}^n\frac{\partial}{\partial V(x_i)}\prod_{i=0}^{m-1}\frac{\partial}{\partial \Psi_R(\tilde{x}_{m-i})}F_{R,g}.
\end{align}
\item[NS-R 5] add the following terms to the right hand side of \eqref{NS2}
\begin{equation}
-\frac{1}{x}\sum_{i=1}^n\frac{d}{dx_i}W_{g,n|m}(J|K)-\frac{1}{x}\sum_{j=1}^m(-1)^{j-1}\frac{d}{d\tilde x_j}W_{g,n|m}(J|K),
\end{equation}
\item[NS-R 6] replace as follows in \eqref{NS4}
\begin{equation}
\frac{1}{x-x_i}\mapsto \frac{1}{x-x_i}\frac{\sqrt{x_i}}{\sqrt{x}},\;\;\;\;\frac{1}{x-\tilde x_j}\mapsto \frac{1}{x-\tilde x_j}\frac{\sqrt{\tilde x_j}}{\sqrt{x}}
\end{equation}
\end{description}
Note that \eqref{R1}  is a consequence of the following result instead of \eqref{NS1}:
\begin{equation}
\frac{\partial}{\partial \Psi_R(x_2)}\Psi_R(x_1)=\frac12\frac{x_1+x_2}{x_1-x_2}\frac{1}{\sqrt{x_1}\sqrt{x_2}}.
\end{equation}

 Crucial differences from a global point of view \cite{O} are not only the appearances of $\sqrt{x}$, but also that $Q^{R,BB}_{g,n+1|m}(x,J|K)$ has a simple pole at $x\rightarrow0$ unlike the $Q^{NS,BB}_{g,n+1|m}(x,J|K)$. As explained in \cite{O}, this originates from the fact that $L_{-1}Z_R\neq0$. However, as long as we stick to a local description in a neighbourhood far from $x\to0$, we do not have to worry about the pole. This is one of advantages of super topological recurison -- one can treat both the Neveu-Schwarz and Ramond sector in the same footing.

\subsubsection{Local Super Spectral Curve and Abstract Super Loop Equations}
Let us find the local super spectral curve for the Ramond sector. It is proven  in \cite{O} that the (global) spectral curve is given by
\begin{equation}
xy^2-(x-1)(M(x))^2=0.
\end{equation}
If we choose the parametrization as
\begin{equation}
x=\frac{1}{1-u^2},\;\;\;\;y=u\,M(x(u)),
\end{equation}
then there are two ramification points $u=0,\infty$. Furthermore, \cite{O} has shown that there is no contribution to the recursion from the irregular ramification point at $u=\infty$ thanks to a supersymmetric cancellation. Hence, we just focus on the regular ramification point $u=0$. This is critical  because $x=0$ at the irregular ramification point $u=\infty$, which is exactly where we would like to avoid because  $Q^{R,BB}_{g,n+1|m}(x,J|K)$ has a simple pole at $x=0$. The supersymmetric cancellation nicely saves us from this issue.

We introduce a local coordinate $z$ in the neighbourhood of $u=0$ by
\begin{equation}
x=1+\frac{z^2}{2}=\frac{1}{1-u^2},
\end{equation}
and extend this patch $\mathbb{C}$ to $\mathbb{C}^{1|1}$. Similar to the Neveu-schwarz sector, we define
\begin{equation}
\omega_{0,1|0}(z|):=M(x)y(z)dx(z)=\frac12\left(W_{0,1|0}^{(0)}(z|)-V'(x(z))\right)dx(z),\label{Rw010}
\end{equation}
\begin{equation}
\omega_{0,2|0}(z_1,z_2|):=\frac12W^{(0)}_{0,2|0}(z_1,z_2|)dx(z_1)dx(z_2)+\frac{dx(z_1)dx(z_2)}{(x(z_1)-x(z_2))^2}=\frac{du(z_1)du(z_2)}{(u(z_1)-u(z_2))^2},
\end{equation}
where they should be understood as formal expansions in $z,z_1,z_2$. The explicit form of $\omega_{0,0|2}(|z_1,z_2)$ was derived in \cite{O} from a global point of view. In our local description, we get
\begin{align}
\omega_{0,0|2}(|z_1,z_2):&=\left(\frac12W^{(0)}_{0,2|0}(|z_1,z_2)-\frac12\frac{x_1+x_2}{x_1-x_2}\frac{1}{\sqrt{x_1}\sqrt{x_2}}\right)\Theta_1\Theta_2\nonumber\\
&=-\frac{(u(z_1)+u(z_2))(1-u(z_1)u(z_2))}{4u(z_1)u(z_2)(u(z_1)-u(z_2))}\left(1+\frac{z_1^2}{2}\right)^{-\frac12}\left(1+\frac{z_2^2}{2}\right)^{-\frac12}\Theta_1\Theta_2,
\end{align}
where this should be understood as a formal expansion in $z_1,z_2$. These $(\omega_{0,1|0}, \omega_{0,2|0}, \omega_{0,0|2})$ provide the defining data of a local super spectral curve for the Ramond sector. Note that instead of \eqref{NSw002} we now have
\begin{equation}
\omega_{0,0|2}(|z,z_1)+\omega_{0,0|2}(|\sigma(z),z_1)=-\frac12\frac{x+x_1}{x-x_1}\frac{1}{\sqrt{x}\sqrt{x_1}}\Theta_1\Theta_2.\label{R2}
\end{equation}

Starting with this local super spectral curve, we can go through procedure (1) - (5) listed in Section~\ref{sec:NS} with only one change, use \eqref{R2} instead of \eqref{NSw002}, and we obtain the abstract super loop equations for the Ramond sector. Moreover, it is shown in \cite{O} that correlation functions respect the polarization. This completes the proof of Proposition~\ref{prop:R}.

\subsection{Proof of Proposition~\ref{prop:truncation}}\label{sec:truncation}
Let us first show that
\begin{equation}
\omega_{0,0|4}=\omega_{0,1|4}=0,\label{truncation1}
\end{equation}
for any local super spectral curve. This can be easily shown by counting the degree of poles. Since this is trivial if $\epsilon=1$,  we focus on local super spectral curves with $\epsilon=3$ and we normalize $\tau_3=1$. Given a regular local super spectral curve, we find
\begin{align}
\omega_{0,3|0}(z_1,z_2,z_3|)&=-d\xi_{-1}(z_1)d\xi_{-1}(z_2)d\xi_{-1}(z_3),\\
\omega_{0,1|2}(z_1|u_1,u_2)&=-\frac12d\xi_{-1}(z_1)\Bigl(\eta_{-2}(u_1)\eta_0(u_2)-\eta_0(u_1)\eta_{-2}(u_2)\Bigr).
\end{align}
The degree of poles of the integrand in \eqref{BR} and \eqref{FR} increases by 2 when $\chi=2g+n+2m$ increases by 1 due to their recursion kernels. In particular, $\omega_{0,0|4}$ can only have poles up to the degree of $\eta_{-4}$ but not higher (this can be explicitly verified with \eqref{FR}). Since there is no $(\eta_{-2l-1})_{l\in\mathbb{Z}}$ due to the linear fermionic loop equations, $\omega_{0,0|4}$ should be given by a linear combination of products of $\{\eta_0,\eta_{-2},\eta_{-4}\}$. However, $\{\eta_0,\eta_{-2},\eta_{-4}\}$ is not enough to construct a completely antisymmetric linear combination. This shows that $\omega_{0,0|4}=0$.

For $\omega_{0,1|4}$, one may naively think that it can have poles up to the degree of $\eta_{-6}$. Indeed, by looking at the pole structure of the integrand, the following terms in \eqref{FR} for $\omega_{0,1|4}$ give:
\begin{align}
&\underset{z\rightarrow0}{{\rm Res}}\,K^{FB}(u_1,z,\sigma(z))\Bigl(\omega_{0,0|2}(|z,u_2)\omega_{0,2|2}(z_1,-z|u_3,u_4)\nonumber\\
&\hspace{30mm}+\omega_{0,1|2}(z_1|z,u_2)\omega_{0,1|2}(-z|u_3,u_4)+(z\leftrightarrow-z)\Bigr)\nonumber\\
&= c\,d\xi_{-1}(z_1)\eta_{-6}(u_1)\eta_0(u_2)\Bigl(\eta_{-2}(u_3)\eta_{0}(u_4)-\eta_{0}(u_3)\eta_{-2}(u_4)\Bigr)\nonumber\\
&\hspace{20mm}+\text{terms independent of }\eta_{-6}(u_1),\label{w014}
\end{align}
for some $c\in\mathbb{C}$. However, after complete antisymmetrization, these terms vanish and we get $\omega_{0,1|4}=0$.  Finally, by induction  we can show that $\omega_{0,0|2m}=0$ by \eqref{BR} and  $\omega_{0,1|2m}=0$  by \eqref{FR} for $m\geq2$.


\newpage
\begin{thebibliography}{9999} 

\bibitem{ABCD}
J.~E.~Andersen, G.~Borot, L.~O.~Chekhov and N.~Orantin,
``The ABCD of topological recursion,"
\arxiv{1703.03307}.



\bibitem{AP}
G.~Akemann and J.~C.~Plefka, ``The Chiral Supereigenvalue Model,'' Mod.Phys.Lett. A12, 1745-1758 (1997) \arxiv{hep-th/9705114}.


\bibitem{SL1} 
  L.~Alvarez-Gaume, H.~Itoyama, J.~L.~Manes and A.~Zadra,
  ``Superloop equations and two-dimensional supergravity,''
  Int.\ J.\ Mod.\ Phys.\ A {\bf 7}, 5337 (1992)
  \arxiv{hep-th/9112018}.
  

\bibitem{AlvarezGaume:1992mm} 
  L.~Alvarez-Gaume, K.~Becker, M.~Becker, R.~Emparan and J.~Manes,
  ``Double scaling limit of the superVirasoro constraints,''
  Int.\ J.\ Mod.\ Phys.\ A {\bf 8}, 2297 (1993)
  \arxiv{hep-th/9207096}.



\bibitem{Beckers} 
  K.~Becker and M.~Becker,
  ``Nonperturbative solution of the superVirasoro constraints,''
  Mod.\ Phys.\ Lett.\ A {\bf 8}, 1205 (1993)
  \arxiv{hep-th/9301017}.




\bibitem{BEO} 
  G.~Borot, B.~Eynard and N.~Orantin,
  ``Abstract loop equations, topological recursion and new applications,''
  Commun.\ Num.\ Theor.\ Phys.\  {\bf 09}, 51 (2015)
  doi:10.4310/CNTP.2015.v9.n1.a2
  \arxiv{1303.5808}.


\bibitem{BS} 
  G.~Borot and S.~Shadrin,
  ``Blobbed topological recursion: properties and applications,''
  Math.\ Proc.\ Cambridge Phil.\ Soc.\  {\bf 162}, no. 1, 39 (2017)
  \arxiv{1502.00981}.  



\bibitem{HAS}
G.~Borot, V.~Bouchard, N.~K.~Chidambaram, T.~Creutzig and D.~Noshchenko,
  ``Higher Airy structures, W algebras and topological recursion,''
  \arxiv{1812.08738}.


\bibitem{SAS}
V.~Bouchard, P.~Ciosmak, L.~Hadasz, K.~Osuga, B.~Ruba and P.~Sułkowski,
  ``Super Quantum Airy Structures,''
  \arxiv{1907.08913}.

\bibitem{BE} 
  V.~Bouchard and B.~Eynard,
  ``Think globally, compute locally,''
  JHEP {\bf 1302}, 143 (2013)
  \arxiv{1211.2302}.

 \bibitem{BKMP}
V.~Bouchard, A.~Klemm, M.~Marino and S.~Pasquetti,
  ``Remodeling the B-model,''
  Commun.\ Math.\ Phys.\  {\bf 287}, 117 (2009)
  \arxiv{0709.1453}.

\bibitem{BM}
V.~Bouchard and M.~Mari\~no,
``Hurwitz numbers, matrix models and enumerative geometry,"
in \emph{From Hodge Theory to Integrability and tQFT: $tt*$-geometry}, Proceedings of Symposia in Pure Mathematics, AMS (2008) \arxiv{0709.1458}.


\bibitem{BO} 
  V.~Bouchard and K.~Osuga,
  ``Supereigenvalue Models and Topological Recursion,''
  JHEP {\bf 1804}, 138 (2018)
  [\arxiv{1802.03536}].
  
  
  \bibitem{BG}
E.~Brezin and D.~J.~Gross,
``The External Field Problem in the Large N Limit of QCD,''
Phys. Lett. B \textbf{97}, 120-124 (1980)
  
\bibitem{BEM} 
  A.~Brini, B.~Eynard and M.~Marino,
  ``Torus knots and mirror symmetry,''
  Annales Henri Poincare {\bf 13}, 1873 (2012)
  \arxiv{1105.2012}.

\bibitem{SJT1} 
  A.~H.~Chamseddine and D.~Wyler,
  ``Gauge Theory of Topological Gravity in (1+1)-Dimensions,''
  Phys.\ Lett.\ B {\bf 228}, 75 (1989).
  
  
\bibitem{CEO} 
  L.~Chekhov, B.~Eynard and N.~Orantin,
  ``Free energy topological expansion for the 2-matrix model,''
  JHEP {\bf 0612}, 053 (2006)
  \arxiv{math-ph/0603003}.
  
  \bibitem{Chen:2020hml}
Y.~Chen, R.~Wang, K.~Wu and W.~Z.~Zhao,
``Correlators in the supereigenvalue model in the Ramond sector,''
\arxiv{2006.11013}.

\bibitem{C1}    
  P.~Ciosmak, L.~Hadasz, M.~Manabe and P.~Sulkowski,
  ``Super-quantum curves from super-eigenvalue models,''
  JHEP {\bf 1610}, 044 (2016)
  \arxiv{1608.02596}.
  
\bibitem{C2} 
  P.~Ciosmak, L.~Hadasz, M.~Manabe and P.~Sulkowski,
  ``Singular vector structure of quantum curves,''
  \arxiv{1711.08031}.



  \bibitem{C} 
  P.~Ciosmak, L.~Hadasz, Z.~Jaskolski, M.~Manabe and P.~Sulkowski,
  ``From CFT to Ramond super-quantum curves,''
  JHEP {\bf 1805}, 133 (2018)
  \arxiv{1712.07354}.

\bibitem{Dijkgraaf}
  R.~Dijkgraaf, H.~L.~Verlinde and E.~P.~Verlinde,
  ``Loop equations and Virasoro constraints in nonperturbative 2-D quantum gravity,''
  Nucl.\ Phys.\ B {\bf 348} (1991) 435.


\bibitem{SL2} 
  J.~Distler, Z.~Hlousek and H.~Kawai,
  ``Superliouville Theory as a Two-Dimensional, Superconformal Supergravity Theory,''
  Int.\ J.\ Mod.\ Phys.\ A {\bf 5}, 391 (1990).

\bibitem{Bessel}
  N.~Do and P.~Norbury,
  ``Topological Recursion on the Bessel Curve,''
  Commun.\ Num.\ Theor.\ Phys.\  {\bf 12}, 53 (2018)
  \arxiv{1608.02781}.




 
 
 

 \bibitem{DBOSS}
P. Dunin-Barkowski, N. Orantin, S. Shadrin and L. Spitz, ``Identification of the Givental formula with the spectral curve topological recursion procedure," Comm. Math. Phys. {\bf 328} 2, 669--700 (2014) \arxiv{1211.4021}.


 \bibitem{E1} 
  B.~Eynard,
  ``Intersection numbers of spectral curves,''
  Communications in Number Theory and Physics (2014), Volume 8,
  Number 3
  \arxiv{1104.0176}.
  
  
 \bibitem{E2} 
  B.~Eynard,
  ``Invariants of spectral curves and intersection theory of moduli spaces of complex curves,''
  Commun.\ Num.\ Theor.\ Phys.\  {\bf 8}, 541 (2014)
  doi:10.4310/CNTP.2014.v8.n3.a4
  \arxiv{1110.2949}.
  
 
\bibitem{EO}
B.~Eynard and N.~Orantin.
``Invariants of algebraic curves and topological expansion,"
Commun. Number Theory Phys., 1(2):347--452 (2007) \arxiv{math-ph/0702045}.

\bibitem{EO2}
B.~Eynard and N.~Orantin.
``Topological recursion in random matrices and enumerative geometry,"
J. Phys. A: Mathematical and Theoretical, 42(29) (2009) \arxiv{0811.3531}.

\bibitem{EO3}
B.~Eynard and N.~Orantin,
``Computation of open Gromov-Witten invariants for toric Calabi-Yau 3-folds by topological recursion, a proof of the BKMP conjecture,"  \arxiv{1205.1103}.


 \bibitem{EMS}
 B.~Eynard, M.~Mulase and B.~Safnuk, ``The Laplace transform of the cut-and-join equation and the Bouchard-Marino conjecture on Hurwitz numbers," Publications of the Research Institute for Mathematical Sciences 47, 629--670 (2011) \arxiv{0907.5224}.
  
  
\bibitem{FLZ2}
B.~Fang, C.~C.~M.~Liu and Z.~Zong, ``On the Remodeling Conjecture for Toric Calabi-Yau 3-Orbifolds," 	\arxiv{1604.07123}.

\bibitem{FLZ3}
B.~Fang, C.~C.~M.~Liu and Z.~Zong, ``The SYZ mirror symmetry and the BKMP remodeling conjecture," 	\arxiv{1607.06935}.

\bibitem{SKdV} 
  J.~M.~Figueroa-O'Farrill and S.~Stanciu,
  ``On a new supersymmetric KdV hierarchy in 2-d quantum supergravity,''
  Phys.\ Lett.\ B {\bf 316}, 282 (1993)
  \arxiv{hep-th/9302057}.
  
  
\bibitem{SJT2} 
  T.~Fukuyama and K.~Kamimura,
  ``Gauge Theory of Two-dimensional Gravity,''
  Phys.\ Lett.\  {\bf 160B}, 259 (1985).

\bibitem{GW}
D.~Gross and E.~Witten,
``Possible Third Order Phase Transition in the Large N Lattice Gauge Theory,''
Phys. Rev. D \textbf{21}, 446-453 (1980)

\bibitem{GJKS} 
  J.~Gu, H.~Jockers, A.~Klemm and M.~Soroush,
  ``Knot Invariants from Topological Recursion on Augmentation Varieties,''
  Commun.\ Math.\ Phys.\  {\bf 336}, no. 2, 987 (2015)
  \arxiv{1401.5095}.

\bibitem{Itoyama} 
  H.~Itoyama,
  ``Integrable superhierarchy of discretized 2-d supergravity,''
  Phys.\ Lett.\ B {\bf 299}, 64 (1993)
  \arxiv{hep-th/9206091}.

\bibitem{K}
M.~Kontsevich,
``Intersection theory on the moduli space of curves and the matrix Airy function,''
Commun. Math. Phys. \textbf{147}, 1-23 (1992)

\bibitem{KS} 
  M.~Kontsevich and Y.~Soibelman,
  ``Airy structures and symplectic geometry of topological recursion,''
  \arxiv{1701.09137}.

%
%
\bibitem{Ma}
M.~Mari\~no, ``Open string amplitudes and large order behavior in topological string theory," JHEP {\bf 0803} 060 (2008) \arxiv{hep-th/0612127}

\bibitem{McArthur} 
  I.~N.~McArthur,
  ``The Partition function for the supersymmetric Eigenvalue model,''
  Mod.\ Phys.\ Lett.\ A {\bf 8}, 3355 (1993).


\bibitem{Mirzakhani1}
M.~Mirzakhani,
``Simple geodesics and Weil-Petersson volumes of moduli spaces of bordered Riemann surfaces,''
Invent. Math. \textbf{167}, no.1, 179-222 (2006)

\bibitem{Mirzakhani2}
M.~Mirzakhani,
``Weil-Petersson volumes and intersection theory on the moduli space of curves,''
J. Am. Math. Soc. \textbf{20}, no.01, 1-24 (2007)


\bibitem{SJT3} 
  D.~Montano, K.~Aoki and J.~Sonnenschein,
  ``Topological Supergravity in Two-dimensions,''
  Phys.\ Lett.\ B {\bf 247}, 64 (1990).
  
  
  \bibitem{N}
P.~Norbury,
``Enumerative geometry via the moduli space of super Riemann surfaces,''
\arxiv{2005.04378}.
  
\bibitem{O} 
  K.~Osuga,
  ``Topological Recursion in the Ramond Sector,''
  JHEP {\bf 1910}, 286 (2019)
  doi:10.1007/JHEP10(2019)286
  \arxiv{1909.08551}.



\bibitem{Plefka1}
J.~C.~Plefka, ``Iterative Solution of the Supereigenvalue Model,'' Nucl.Phys. B444, 333-352 (1995) \arxiv{hep-th/9501120}.


\bibitem{Plefka2}
J.~C.~Plefka, ``The Supereigenvalue Model in the Double-Scaling Limit,'' Nucl.Phys. B448, 355-372 (1995) \arxiv{hep-th/9504089}.

%


   
\bibitem{SW} 
  D.~Stanford and E.~Witten,
  ``JT Gravity and the Ensembles of Random Matrix Theory,''
  \arxiv{1907.03363}.


\bibitem{W}
E.~Witten,
``Two-dimensional gravity and intersection theory on moduli space,''
Surveys Diff. Geom. \textbf{1}, 243-310 (1991)





%
%


\bibitem{W5}
E.~Witten,
``Volumes and Random Matrices,''
\arxiv{2004.05183}.

\bibitem{SL3} 
  A.~Zadra and E.~Abdalla,
  ``Noncritical superstrings: A Comparison between continuum and discrete approaches,''
  Nucl.\ Phys.\ B {\bf 432}, 163 (1994)
  \arxiv{hep-th/9402083}.


\end{thebibliography}
\end{document}